\documentclass[12pt]{article}
\usepackage{techart}
\usepackage{pmat}

\renewcommand\vec[1]{\boldsymbol{#1}}
\renewcommand\mat[1]{\boldsymbol{#1}}

\newcommand\transpose[1]{\ensuremath{{#1}^T}}

\newcommand\x{x}
\newcommand\y{y}
\newcommand\z{z}

\newcommand\Y{Y}
\newcommand\Z{Z}

\newcommand\vx{\vec{x}}
\newcommand\vy{\vec{\y}}
\newcommand\vY{\vec{\Y}}

\newcommand\M{\mathcal{M}}

\newcommand\mH{\mat{H}}

\newcommand\anchor{\ensuremath{\theta}}
\newcommand\Anchor{\ensuremath{\Theta}}

\newcommand\err{\epsilon}
\newcommand\gauss{\phi}

\newcommand\vmu{\vec{\mu}}
\newcommand\mSigma{\mat{\Sigma}}

\begin{document}
\title{Adaptive Anchored Inversion for Gaussian\\ Random Fields
Using Nonlinear Data}
\author{Zepu Zhang}
\date{October 30, 2011}
\maketitle

\barefoot{zepu.zhang@gmail.com\\
Department of Mathematics and Statistics, University of Alaska,
Fairbanks, Alaska, USA.\\[2ex]
Published as \textit{Inverse Problems}
\textbf{27}(2011) 125011. doi:10.1088/0266-5611/27/12/125011}

\vspace*{-1cm}

\begin{abstract}
In a broad and fundamental type of ``inverse problems'' in science,
one infers a spatially distributed physical attribute based on
observations of processes that are controlled by the spatial attribute
in question.
The data-generating field processes,
known as ``forward processes'', are usually
nonlinear with respect to the spatial attribute,
and are often defined non-analytically by a numerical model.
The data often contain a large number of elements with
significant inter-correlation.
We propose a general statistical method to tackle this problem.
The method is centered on a parameterization device called ``anchors''
and an iterative algorithm for deriving the distribution of anchors
conditional on the observed data.
The algorithm draws upon
techniques of importance sampling and
multivariate kernel density estimation with weighted samples.
Anchors are selected automatically;
the selection evolves in iterations in a way
that is tailored to important features of the attribute field.
The method and the algorithm
are general with respect to the scientific nature and
technical details of the forward processes.
Conceptual and technical components
render the method in contrast to
standard approaches that are based on regularization or optimization.
Some important features of the proposed method
are demonstrated by examples
from the earth sciences,
including groundwater flow, rainfall-runoff, and seismic tomography.

\end{abstract}

\section{Introduction}
\label{sec:intro}

Consider a spatial Gaussian process (also called a Gaussian random field)
$\Y(\x)$,
where $\x \in \mathbb{D} \subset \mathbb{R}^3$,
$\mathbb{D}$ being the model domain,
and $\Y \in \mathbb{R}$.
Define a \emph{nonlinear} functional of this field:
$\M\colon \Y(\mathbb{D}) \to \Z$,
where $\Z \in \mathbb{R}^{n_\z}$ is a vector of length $n_\z \ge 1$.
Suppose we have observed the value of this functional,
say the vector value is $\z$.
Conditional on this observation,
how can we characterize the field $\Y(\x)$?

This is an abstraction of statistical approaches to a wide range of
scientific questions.
In these questions,
$\Y(\x)$ is a ``spatially-distributed'' physical attribute.
The functional $\M$, known as the ``forward model'' or ``forward
process'',
is typically embodied in a deterministic, numerical algorithm that
takes the field $\Y(\mathbb{D})$ as input
and outputs vector-valued result $\Z$.
In nature, the forward model is a physical process whose outcome
is dictated by the attribute $\Y$ in the entire domain $\mathbb{D}$.
The task is, conditional on the observed outcome of the forward model,
how can we characterize (or ``back out'') the attribute field $\Y$?
Since we are using the outcome of the forward model
to infer the quantity $\Y$ that dictates the outcome of the forward
model,
this task is known as an ``inverse problem''.

A few examples will make the setting clear.

\emph{Example 1}.
\citet{Yeh:1986:RPI} reviews techniques for the groundwater
inverse problem, in which the spatial attribute of interest is hydraulic
conductivity of the subsurface medium.
Conductivity controls groundwater flow
which is governed by well-understood partial differential equations.
Scientists or engineers make observations of groundwater movement
to get head (\ie pressure) and flux at selected locations and times,
and attempt to use these data to infer the spatially distributed
conductivity.

\emph{Example 2}.
\citet{Bube:1983:ODI} study a 1-D problem in exploration seismology.
In this setting,
an impulsive or vibrating load applied at the ground surface launches
elastic waves into the earth's interior.
Part of the wave energy is reflected by the medium and reaches the ground
surface, where it is monitored at many time instants.
Wave propagation in the subsurface column
is described by a hyperbolic equation system
with mechanic attributes of the elastic medium,
including density and Lam\'e constants,
as input parameters.
The task is to recover the mechanic attributes,
ideally everywhere in the subsurface column,
using surface measurements of wave propagation
including pressure and particle velocity.

\emph{Example 3}.
\citet{Newsam:1988:IPA} deduce the spatial
distribution of $\text{CO}_2$ sources (including sinks) on the surface of the globe,
given time-series observations of surface concentrations of
$\text{CO}_2$ at some 20 locations around the globe.
The $\text{CO}_2$ source here is treated as a spatial attribute defined
for every location on the ground surface.
The surface $\text{CO}_2$ concentration around the globe
is the result of re-distribution of the sources
via a transport process, governed by a diffusion equation.

These examples have three components in common:
a spatial attribute of interest
(hydraulic conductivity of flow media;
mechanic properties of elastic media;
source/sink of CO$_2$),
a forward process
(groundwater flow; wave propagation; atmospheric transport),
and observed outcomes of the forward process
(head and flux; pressure and particle velocity; CO$_2$ concentration).
The attribute of interest varies in space and is possibly highly
heterogeneous; in some scientific literature it is called a
``(spatially) distributed parameter''
\citep{Kravaris:1985:IPD, Beven:1993:PRU, Nakagiri:1993:RJW}.
The forward process is represented by a numerical model,
which takes the spatial attribute as the main input,
along with boundary and initial conditions.
Usually, the need for this inversion
arises because the attribute $\Y(\x)$ is needed for modeling
(or simulating, or predicting) some other process of interest.

In concept,
the unknown, $\Y(\mathbb{D})$, has infinite dimensions.
In real-world applications,
it is almost always defined on a finite grid that discretizes the model
domain.
In either case,
the unknown $\Y(\mathbb{D})$ has many more dimensions than the known $\Z$.
This makes the inverse problem a severely under-determined one.
It has become an established strategy to formulate the inverse problem
in a statistical way
\citep{Kirsch:1996:IMT, Kaipio:2005:SCI}, that is,
take $\Y(\x)$ as a random field and try to obtain its
\emph{conditional distribution} given the observation $\z$,
\ie $p\bigl(\Y(\mathbb{D}) \given \z\bigr)$.
Typically,
this conditional distribution is represented empirically by
a large number of \emph{realizations} (\ie simulations)
of the random field.
In this context,
Bayesian approaches and Markov chain Monte Carlo computations
have been widely used (see section~\ref{sec:anchor-motivation}).

In this study,
we propose a method for this inverse problem.
The proposal centers on a parameterization device called ``anchors'',
hence the name of the method---``anchored inversion''.
Let the anchor parameters be denoted by $\anchor$.
With anchor parameterization,
\emph{we solve a reduced problem},
namely, $p(\anchor \given \z)$ rather than
$p\bigl(\Y(\mathbb{D}) \given \z\bigr)$.
The dimension of $\anchor$ is much lower than that of
$\Y(\mathbb{D})$ (the latter could be infinite),
and the dimension of $\anchor$ is under the user's control.
This dimension reduction brings advantages in computation as well as
conceptualization.

The proposed method is a fundamental departure from some existing
approaches.
In section~\ref{sec:anchor-motivation},
we identify some challenges faced by existing methods,
which motivate the concept of ``anchor''.
Anchor parameterization is formally described
in section~\ref{sec:anchor-parameterization}.
Section~\ref{sec:inference} presents an iterative algorithm
that approximates the posterior distribution of
the chosen anchor parameter ($\anchor$) by a normal mixture.
In section~\ref{sec:choice-of-anchors} we describe
a strategy for choosing anchor parameters
automatically and adaptively as the iterative algorithm proceeds.
This efficient strategy makes anchor parameters evolve
in iterations, adapted to the available data,
complexity of the $Y$ field, and the computational effort so far invested.
This anchor-selection procedure is integrated into the algorithm
described in section~\ref{sec:inference}.

Section~\ref{sec:extensions}
describes two extensions to the basic framework presented in earlier
sections.
The first extension, geostatistical parameterization,
is expected to be used in most applications.
The second extension is used whenever the type of data described there
are available.
In section~\ref{sec:examples},
the method is applied to three scientific questions using synthetic
data.
Section~\ref{sec:conclusion} concludes with a summary and highlights.



We emphasize ``nonlinear'' in this study,
because treatment of ``linear'' forward models is relatively mature.
A mainstay of the approaches
is the Kalman filter and variants \citep{Welch:1995:IKF}.
However, the vast majority of forward processes in scientific
applications are nonlinear,
as demonstrated in section~\ref{sec:examples}.
Linear data, if available,
are accommodated as described in
section~\ref{sec:linear-data}.

Although $\Y$ is ``point-referenced'' in concept,
in practice the model domain is discretized (or
aggregated) into a numerical grid.
This discretization is assumed throughout,
and it enables the use of matrix notations instead of integrals.
We shall use $\x$ and $\Y$ as generic symbols for
location and the spatial variable, and use
$\vx$ and $\vY$ for the finite-length $\x$ and $\Y$ vectors
corresponding to the numerical grid.
Specific values of $\Y$ and $\vY$ are denoted by $\y$ and $\vy$,
respectively.
Throughout, the symbol $\gauss(...)$ denotes a normal density or
likelihood function.
We use $p(\cdot)$ as a generic symbol for probability density function
and $p(\cdot\given \cdot)$ for conditional density.
When we have a specific density, say obtained by approximation,
we use a specific symbol, for example $f(\cdot)$.


\section{Motivation and rationale for the concept of anchor}
\label{sec:anchor-motivation}

Discretization of the model domain, along with other sources,
introduces model error in the simulation of the forward process, $\M$.
In addition there may be measurement error in $\z$.
Suppose the model and measurement errors are additive, we can
write
\begin{equation}\label{eq:additive-error}
\z = \M(\vy) + \err
,
\end{equation}
where
$\z$ is the forward data,
$\M$ is the forward model
(\ie a functional of the discretized field $\vy$),
and $\err$ is combined model and measurement error.

A typical Bayesian approach takes $\vY$ as the parameter vector
and seeks its posterior distribution given the observation $\z$.
Suppose the density of the error $\err$ is known to be
$f(\err)$,
then the likelihood function is
\begin{equation}\label{eq:additive-likelihood}
p(\z \given \vy)
= f\bigl(\z - \M(\vy)\bigr)
.
\end{equation}
Upon specification of a prior $\pi(\vy)$,
the posterior is
$p(\vy \given \z) \propto \pi(\vy)\, p(\z \given \vy)$.
This high-dimensional posterior distribution is usually
represented by samples from it via
Markov chain Monte Carlo (MCMC) or related methods.
There exist a large number of studies along this line,
\eg~\citet{vanLeeuwen:1996:DAI, Lee:2002:MRF, Sambridge:2002:MCM,
Kaipio:2005:SCI, Tarantola:2005:IPT}.
If the error is multiplicative with known distribution,
manipulations may be possible in simple cases;
see \citet[p.~58]{Kaipio:2005:SCI} for an example.
The construct above is also the cornerstone for
a large class of approaches using
``regularization'' and optimization
\citep{Tikhonov:1977:SIP, Kravaris:1985:IPD,
Engl:2000:RIP, Tenorio:2001:SRI, Doherty:2003:GWM}.

We recognize three difficulties in this approach, namely,
(1) that the same $\vY$ is both the model parameter and the input
to the forward model creates conflicting requirements;
(2) the formulation is centered at ``error'', $\err$,
ruling out situations where error is nonexistent or negligible;
(3) distribution of the error $\err$,
especially the model error, is hard to specify.
These points are elaborated below.
Additional comments can be found in \citet{OSullivan:1986:SPI}.

\begin{enumerate}
\item[(1)]
By taking the entire model grid $\vY$ as the parameter vector,
the dimension of the parameter space is tied to the spatial resolution of the
numerical implementation of the forward model $\M$.
This creates conflicts between parameter identifiability and accuracy in the
numerical forward model $\M$,
which is the essential connection between the model parameter and data.
One side of this paradox calls for a coarse grid
(for better identification of the model parameter)
whereas the other side prefers a fine grid
(for more accurate simulation of the forward process).

\item[(2)]
The only randomness in the parameter-data link
in this formulation arises from the model and measurement
errors, $\err$, because the forward model $\M$ is deterministic.
While in reality error usually exists,
in synthetic or theoretical studies it does not have to.
Clearly, the Bayesian approach centered at~(\ref{eq:additive-likelihood})
does not apply if error is nonexistent.
On the other hand,
even if the magnitude of the error is not quite negligible,
one may elect to ignore errors (because, for example,
knowledge about the error's distribution is very limited and unreliable)
and still have a meaningful inverse problem
of inferring $\vY$ from $\z$.
However, the formulation based on~(\ref{eq:additive-error})
is not applicable in this situation.

\item[(3)]
Reliable knowledge of the error distribution,
$f(\err)$, is often unavailable.
A major complication here is model error.
While one may have a decent knowledge of the \emph{measurement error} based on
instrument specifications, \emph{model error} is much more elusive.
There are numerous sources for model error, including
spatial and temporal discretization,
inaccurate boundary and initial conditions,
relevant physics that are omitted from consideration,
and so on.
Another notable error stems from the so-called incommensurability
\citep{Ginn:1990:IMS},
\ie the fact that what is computed by the model
and what is measured in the field are not exactly the same quantity.
A common cause of incommensurability is
discrepancy in the spatiotemporal resolution (or scale).
Taken as a random vector,
the model error usually has inter-correlated components.
To further complicate the matter,
the distribution of this error vector may depend
on the input field $\vy$.

\citet{Sambridge:2002:MCM} point out that it is especially difficult to
know the statistics of all errors when
``the theoretical predictions from a model involve approximations''.
However,
all the sources of model error listed above essentially result in
approximations.

The fact that $\err$ contains model error
has been emphasized by \citet{Scales:2001:PIU, Lee:2002:MRF,
Higdon:2003:MCM}.
At this point it must be recognized that model error may not be smaller
than measurement error,
hence it is not feasible to assume that measurement error
dominates model error and makes the latter negligible.
\citet{Tarantola:1982:IPQ} give a telling example:
``In seismology, the theoretical error made by solving the forward
travel time problem is often one order of magnitude larger than the
experimental error of reading the arrival time on a seismogram.''
Also see \citet{Schoups:2010:CAE}.
\end{enumerate}

We present an alternative approach
that has the potential to alleviate these difficulties.
The first intuition comes from the difficulty~(1) above.
Realizing that it is largely hopeless to aim for a sharp resolution of, say,
a $100 \times 100$ field $\vY$ using a data vector $\z$ of length 100,
we ask,
``Instead of taking on the entire field $\vY$,
what about aiming for a sharper resolution of, say, 50
`characteristic values' of $\vY$ using the data $\z$, and then
parameterizing the field with these 50 characteristic values?''

The answer is to reduce the model parameter vector and separate it from
the numerical grid.
Specifically,
we take certain linear functionals of the field as such characteristic
values and call them ``anchors'', denoted by $\anchor$:
\begin{equation}\label{eq:anchor-def}
\anchor = \mH\vY
,
\end{equation}
where $\mH$ is a $n_\anchor \times n_{\vy}$ matrix
of rank $n_\anchor$, satisfying $n_\anchor < n_{\vy}$.
Subsequently, our model inference involves deriving
$p(\anchor \given \z)$,
and parameterizing the field $\vY$ by $\anchor$
via $p(\vy \given \anchor)$.
In a sense, we have proposed to solve a different problem,
a ``reduced'' one,
than that tackled by the standard approach
(which derives $p(\vy \given \z)$).
In the reduced problem,
the anchors act as ``middleware'' between $\z$ and $\vY$;
the most important information in $\z$ about $\vY$
is ``transferred'' into $\anchor$,
in the form of $p(\anchor\given \z)$, instead of into
$\vY$ directly.
We shall use the name ``anchored inversion'' for the proposed method.
The key elements of this method are
the anchor definition represented by the matrix $\mH$,
the anchor parameterization for the field, \ie $p(\vy \given \anchor)$,
and derivation of the posterior $p(\anchor \given \z)$.
These issued are discussed in the sections to come.

It is apparent that the anchor concept de-couples
the dimension of the model parameter (which is now $\anchor$)
from that of the numerical grid.
The former is under our control,
and as such can be vastly lower than the latter.
Whereas the dimension of $\vY$ may be dictated by physics,
required resolution and accuracy of the forward model,
as well as subsequent application needs for the
inferred attribute field,
the dimension of $\anchor$ is determined mainly by statistical and
computational concerns.
The user has the freedom to seek a trade-off,
via the choice of anchors,
between feasible derivation of $p(\anchor \given \z)$
(calling for a shorter $\anchor$)
and sufficient parameterization of the
field by $p(\vy \given \anchor)$ (calling for a longer $\anchor$).

This spirit of dimension reduction is shared by an active line of
research using ``convolutions''
\citep{Higdon:2002:SST, Higdon:2003:MCM, Short:2010:PVC}.
The convolution-based methods differ from anchored inversion
in many ways.
For one thing,
the former conducts computation in a MCMC framework,
whereas the latter is ``incompatible'' with MCMC, as addressed next.

The second difficulty in the existing approach,
\ie errors must be present for the formulation to be applicable,
is eliminated.
In the existing approach,
the parameter-data connection is
\[
\vy
\xrightarrow{\;\M(\vy),\, \err\;} \z
.
\]
In contrast,
the connection in anchored inversion is
\begin{equation}\label{eq:anchor-data-randomness}
\anchor
\xrightarrow{\;p(\vy \given \anchor)\;} \vy
\xrightarrow{\;\M(\vy),\, \err\;} \z
.
\end{equation}
The randomness in the parameter-data connection now comes from two sources
(or layers):
the random field $\vy$ conditional on parameter $\anchor$,
and the random error $\err$ conditional on fixed $\vy$.
The algorithm of anchored inversion (see section~\ref{sec:inference})
learns about the statistical relation between the model parameter
$\anchor$ and the data $\z$ (or $\Z$, to be more accurate)
by simulating the data-generation process.
Note that the first layer of randomness is tractable by simulation
because it is due to the \emph{known} parameterization
$p(\vy \given \anchor)$.
The second layer of randomness can be simulated if
one has decent knowledge about the errors.
The ``knowledge'' here does not have to be a closed-form distribution;
it may be, for example, mechanism of the occurrence of errors in
just enough detail such that the (stochastic) mechanism can be embedded
in the forward simulation.

However,
if the dimension of $\anchor$ is much lower than that of $\vy$,
the first layer of randomness could dominate the second,
making an accurate quantification of $\err$ less critical
to accurately simulating the statistical relation between
$\anchor$ and $\z$.
In such situations,
inaccurate but helpful quantification of the errors may be used,
or the errors may be ignored altogether.
Moreover, anchored inversion allows the situation where
the forward model and measurements are error free because,
lacking the second layer of randomness,
the parameter-data relation is still a statistical one due to the first
layer of randomness.
(In fact,
in the case studies presented in section~\ref{sec:examples},
we did not artificially add model or measurement errors to the synthetic
data.)

This leads to the most fundamental distinction
between anchored inversion and MCMC methods:
the former does not make analytical quantification
of the statistical relation between the model parameter and data,
but rather simulates the relation.
In the simulation,
a likelihood function is not known (and not needed);
consequently, a standard application of the Bayes theorem is not
possible.

Related to the third difficulty in the existing approach,
general forms of errors,
be it additive, multiplicative, or arising
in multiple stages of the forward process $\M$,
may be accommodated in the numerical derivation
of the posterior $p(\anchor \given \z)$.
This will be discussed in section~\ref{sec:accommodate-errors}.
Section~\ref{sec:accommodate-errors} also suggests
that it may be possible to apply the proposed method to inverse problems
where the forward model is non-deterministic.

\section{Anchor parameterization}
\label{sec:anchor-parameterization}

We model $\Y(\x)$ as a Gaussian process.
In other words, $\vY(\vx)$ is a (high dimensional) normal random vector
with mean $\vmu$ and covariance matrix $\mSigma$:
\begin{equation}\label{eq:field-prior}
p(\vy\given \vmu, \mSigma) = \gauss\bigl(\vy \given \vmu, \mSigma\bigr)
,
\end{equation}
where $\gauss$ is the normal density function.
With the anchor $\anchor$ defined in~(\ref{eq:anchor-def}),
we have the joint distribution
\[
\begin{pmat}[{}]
    \vY \cr\-
    \anchor\cr
    \end{pmat}
\sim N\Biggl(
        \begin{pmat}[{}]
            \vmu\cr\-
            \mH\vmu\cr
            \end{pmat}
        ,\;
        \begin{pmat}[{|}]
            \mSigma & \mSigma\transpose{\mH}\cr\-
            \mH\mSigma & \mH\mSigma\transpose{\mH}\cr
            \end{pmat}
    \Biggr)
.
\]
Then $\vY$ conditional on $\anchor$ is normal:
\begin{equation}\label{eq:Y-given-anchor}
p(\vy \given \anchor, \vmu, \mSigma)
= \gauss\Bigl(
    \vy
    \biggiven
    \vmu
    + \mSigma\transpose{\mH}
        \bigl(\mH\mSigma\transpose{\mH}\bigr)^{-1}
        \bigl(\anchor - \mH\vmu\bigr)
    ,\;
    \mSigma
    - \mSigma\transpose{\mH}
        \bigl(\mH\mSigma\transpose{\mH}\bigr)^{-1}
        \mH\mSigma
    \Bigr)
.
\end{equation}

The power of the anchor parameterization lies in the basic property of
the normal distribution that leads to the conditional
distribution~(\ref{eq:Y-given-anchor}).
Depending on the choice of anchors (\ie the matrix $\mH$),
this parameterization introduces flexible, and analytically known,
mean and covariance structures into the field vector $\vY$.

We shall speak of ``sampling'' the conditional distribution
$p(\vy \given \anchor, \vmu, \mSigma)$.
In actual implementations,
an explicit sampling of~(\ref{eq:Y-given-anchor}) is often replaced by
a ``conditional simulation'' procedure that consists of two steps:
\\ \strut\hspace{\parindent}%
(a) Sample $\vy^*$ from
$\gauss\bigl(\vy\given \vmu, \mSigma\bigr)$.
\\ \strut\hspace{\parindent}%
(b) Update $\vy^*$ to
$
\vy_* =
\vy^* + \mSigma \transpose{\mH}
    \bigl(\mH \mSigma \transpose{\mH}\bigr)^{-1}
        (\anchor - \mH\vy^*)
$.
\\
It is easily verified that the resultant $\vy_*$
is indeed a random draw from~(\ref{eq:Y-given-anchor}).

A typical application of anchored inversion consists of three
steps:

(a) Define anchors $\anchor$, \ie specify $\mH$.

(b) Derive the distribution of $\anchor$ conditional on data $\z$,
that is, $p(\anchor \given \z, \vmu, \mSigma)$.

(c) Generate a sample of field realizations
using
the posterior $p(\anchor \given \z, \vmu, \mSigma)$ and
the conditional $p(\vy \given \anchor, \vmu, \mSigma)$.

We first present an iterative algorithm in section~\ref{sec:inference}
for approximating $p(\anchor\given \z, \vmu, \mSigma)$,
assuming $\mH$ has been specified.
In section~\ref{sec:choice-of-anchors},
we present a procedure that automatically and adaptively specifies
$\mH$ in the iterations of the algorithm.
Therefore the anchors (\ie $\mH$) are not fixed,
but rather evolve in the iterations.
The third step entails sampling by ``composition''
\citep[sec~3.3.2]{Tanner:1996:TSI}, that is,
sampling $\anchor_*$ from $p(\anchor \given \z, \vmu, \mSigma)$
followed by
sampling $\vy_*$ from $p(\vy \given \anchor_*, \vmu, \mSigma)$.
In these samplings,
the anchor specification and their posterior distribution are
taken from the final iteration of the algorithm.
Because $p(\vy \given \anchor_*, \vmu, \mSigma)$ is normal,
the computational cost of the sampling is low.
Sampling $p(\anchor \given \z, \vmu, \mSigma)$ is also easy because,
as we shall see in section~\ref{sec:inference},
the distribution is approximated by a normal mixture distribution.

In section~\ref{sec:geostat-extension},
we will introduce a small number of geostatistical parameters to
characterize the mean vector $\vmu$ and the covariance matrix $\mSigma$ of the
field.
These geostatistical parameters are regarded as unknown
and are inferred in the iterative algorithm
along with the anchor parameters.

\section{Iterative conditional mixture approximation to the posterior
distribution}
\label{sec:inference}

We now tackle the posterior of the model parameter $\anchor$ given
data $\z$, the mean vector $\vmu$,
and the covariance matrix $\mSigma$ of the field vector $\vY$.
Let $\pi(\anchor)$ be a prior for $\anchor$.
This is specified by
\begin{equation}\label{eq:anchor-prior}
\pi(\anchor)
= p(\anchor \given \vmu, \mSigma)
= \gauss(\anchor\given \mH\vmu, \mH\mSigma\transpose{\mH})
\end{equation}
following (\ref{eq:anchor-def}) and~(\ref{eq:field-prior}).
Noticing the parameter-data connection depicted
in~(\ref{eq:anchor-data-randomness}) and
the generality of the forward model $\M(\vy)$,
the usual route through the Bayes theorem,
$p(\anchor \given \z, \vmu, \mSigma)
\propto \pi(\anchor)\, p(\z \given \anchor, \vmu, \mSigma)$
is not usable here because
the likelihood function $p(\z \given \anchor, \vmu, \mSigma)$ is unknown.
In fact,
it can be difficult to analyze the statistical relation between
$\anchor$ and $\Z$ because $\M$ is typically
defined by a complicated numerical code,
involving many components and steps,
including possibly ad hoc operations.

However,
we can sample $p(\z \given \anchor, \vmu, \mSigma)$
following the ``data-generating'' mechanism
$\anchor \longrightarrow \vy \longrightarrow \z$.
Starting with a particular $\anchor_*$,
this mechanism involves
sampling $\vy_*$ from $\gauss(\vy \given \anchor_*, \vmu, \mSigma)$
and subsequently evaluating $\z_* = \M(\vy_*)$.
If $\anchor_*$ is a random sample from the prior
$\pi(\anchor)$, then $(\anchor_*, \z_*)$ is a random sample from the joint distribution of
$\anchor$ and $\Z$,
namely
$p(\anchor, \z \given \vmu, \mSigma)
= \pi(\anchor)\, p(\z \given \anchor, \vmu, \mSigma)$.

In general, suppose $(\anchor, \Z)$
(conditional on $\vmu$ and $\mSigma$)
has a mixture distribution with density
\[
p(\anchor, \z \given \vmu, \mSigma) = \sum_{i=1}^n w_i\, p_i(\anchor, \z)
,
\]
where $p_i$ is the density of the $i$th mixture component and
the weights $w_i$ sum to unit.
Denote the marginals of $\anchor$ and $\Z$ corresponding to
$p_i(\anchor, \z)$ by
$p_i(\anchor)$ and $p_i(\z)$, respectively,
and write the conditionals
$p_i(\anchor \given \z) = p_i(\anchor, \z) / p_i(\z)$ and
$p_i(\z \given \anchor) = p_i(\anchor, \z) / p_i(\anchor)$.
The density of $\anchor$ conditional on $\z$
as well as $\vmu$ and $\mSigma$ is then
\begin{equation}\label{eq:mixture-conditional}
p(\anchor \given \z, \vmu, \mSigma)
= \frac{p(\anchor, \z)}{\int_{\Anchor} p(\anchor, \z) \diff \anchor}
= \frac{\sum_i w_i\, p_i(\anchor, \z)}
    {\sum_j w_j \int_{\Anchor} p_j(\anchor, \z) \diff \anchor}
= \frac{\sum_i w_i\, p_i(\anchor, \z)}
    {\sum_j w_j\, p_j(\z)}
= \sum_i v_i\, p_i(\anchor \given \z)
\end{equation}
where
$v_i = w_i\,p_i(\z) / \sum_j w_j\, p_j(\z)$.
This shows that the conditional (or posterior)
$p(\anchor \given \z, \vmu, \mSigma)$ is a mixture
of the conditionals $p_i(\anchor \given \z)$.

Take a pause here and think about the situation we are in.
First, it appears possible to obtain random samples of  $(\anchor, \Z)$.
Second, the joint density of $(\anchor, \Z)$
may be approximated by a mixture distribution.
Third, if it is feasible to get the marginal
$p_i(\z)$ and conditional
$p_i(\anchor \given \z)$
from the mixture component $p_i(\anchor, \z)$,
we are on a clear path to $p(\anchor \given \z, \vmu, \mSigma)$,
the ultimate target.

Regarding the last point, $p_i(\z)$ and $p_i(\anchor\given \z)$
are readily available if $p_i(\anchor, \z)$ is normal.
The second point suggests kernel density estimation.
In view of the third point, Gaussian kernels should be used.
As for the first point,
if we do not have random samples of $(\anchor, \Z)$,
we could have ``weighted'' samples by the general
importance sampling technique
\citep{Geweke:1989:BIE, Owen:2000:SEI}.

In fact the algorithm below will show that importance sampling
is critical for building an \emph{iterative} procedure.
Because we are able to sample $(\anchor, \Z)$ as needed
(via simulations),
we sample it in iterations,
each time using a manageable sample size and an ``improved'' proposal
distribution.
After obtaining a reasonably large sample of $(\anchor, \Z)$,
we embark on a Gaussian kernel density estimation
for the joint distribution $p(\anchor, \z \given \vmu, \mSigma)$,
with the real goal being an estimate of the posterior (or conditional)
$p(\anchor \given \z, \vmu, \mSigma)$.
Over iterations,
the proposal distribution improves
not in terms of estimating $p(\anchor, \z \given \vmu, \mSigma)$
but in terms of estimating $p(\anchor \given \z, \vmu, \mSigma)$.

\subsection{Algorithm}
\label{sec:algorithm}

We begin with an initial approximation to the posterior
$p(\anchor \given \z, \vmu, \mSigma)$, denoted by $f^{(0)}(\anchor)$,
which is taken to be a sufficiently diffuse normal distribution.
The first iteration updates $f^{(0)}$ to $f^{(1)}$.
In general,
the $k$th iteration updates the approximation $f^{(k-1)}$
to $f^{(k)}$ as follows.

\begin{enumerate}
\item \emph{Drawing a random sample of the parameters.}

Take $n$ random samples, denoted by
$\anchor_1,\dotsc,\anchor_n$, from $f^{(k-1)}$.
For $i=1,\dotsc,n$,
compute the prior density
$ s_i = \pi(\anchor_i) $
according to~(\ref{eq:anchor-prior}),
and the ``proposal density''
$ t_i = f^{(k-1)}(\anchor_i) $;
let $w_i = \frac{s_i / t_i}{\sum_{j=1}^n s_j/t_j}$.

\item \emph{Evaluating the forward model.}

For $i=1,\dotsc,n$,
sample $\vy_i$ from $p(\vy \given \anchor_i, \vmu, \mSigma)$
according to the density~(\ref{eq:Y-given-anchor}),
and evaluate $\z_i = \M(\vy_i)$.
The sampling of $\vy_i$ is typically accomplished by a geostatistical
simulation algorithm outlined in
section~\ref{sec:anchor-parameterization},
while evaluation of $\z_i$ usually entails
running a numerical forward model with the simulated field $\vy_i$
as input.

\item \emph{Reducing the dimension of the conditioning data
by principal component analysis (PCA).}

Let
$ \mat{Z} = \transpose{
    \begin{pmat}[{||}] \z_1 & \hdots & \z_n\cr \end{pmat}}$,
where $\z_1,\dotsc,\z_n$ are column vectors of length $n_\z$.
By PCA,
we find a $n_\z \times m$ matrix $\mat{A}$,
where $m \le n_\z$,
such that the matrix $\mat{Z} \mat{A}$
``explains'' specified proportions (\eg 99\%)
of the variations in $\mat{Z}$.
Subsequently,
we transform each simulated $\z_i$, $i=1,\dotsc,n$, to
$\transpose{\mat{A}} \z_i$,
and also transform the observation vector $\z$ similarly.
This transformation reduces the dimension of $\Z$
from $n_\z$ to $m$.
To keep the notation simple,
we shall continue to use the symbol $\z$ for this transformed variable.

\item \emph{Estimating the joint density of $(\text{\tt parameter,
data})$.}

Based on the \emph{weighted} sample
$\{(\anchor_1, \z_1),\dotsc, (\anchor_n, \z_n)\}$
with weights $w_i$,
approximate the density function of the random vector
$(\anchor, \Z)$ by a multivariate normal mixture:
\begin{equation}\label{eq:mixture-normal}
p(\anchor, \z)
\approx \sum_{i=1}^n w_i\, p_i(\anchor,\z)
= \sum_{i=1}^n w_i\, \gauss(\anchor, \z\given \vec{m}_i, \mat{V}_i),
\end{equation}
where $\vec{m}_i$ and $\mat{V}_i$ are the mean vector and covariance
matrix of the $i$th mixture component.

We use a normal-kernel density estimator with two tuning parameters:
one is bandwidth, which is a standard tuning parameter;
the other is a localization parameter,
which specifies how large a fraction of the entire sample
is used to calculate the covariance matrix of each mixture component.
These two tuning parameters are determined simultaneously by a
likelihood-based optimization procedure.
Because this step is a completely modular component
in the entire algorithm
(and alternative normal-kernel density estimators may be used
without any change to other parts of the algorithm),
we refer the reader to \citet{Zhang:2012:IPI} for details.

\item \emph{Conditioning on the data.}

For $i=1,\dotsc,n$,
partition $\mat{V}_i$ as
$
\begin{pmat}[{|}]
\mat{V}_{\anchor\anchor i} & \mat{V}_{\anchor\z i}\cr\-
\mat{V}_{\z\anchor i} & \mat{V}_{\z\z i}\cr
\end{pmat}
,
$
where $\mat{V}_{\anchor\anchor i}$ is the covariance matrix of $\anchor$,
$\mat{V}_{\z\z i}$ is the covariance matrix of $\Z$,
and $\mat{V}_{\anchor\z i}$ as well as $\mat{V}_{\z\anchor i}$ are cross-covariance matrices.
Similarly,
decompose $\vec{m}_i$ into $\vec{m}_{\anchor i}$ and $\vec{m}_{\z i}$,
representing the $\anchor$ part and $\Z$ part, respectively.
Use the notations
$p_i(\anchor)$, $p_i(\z)$, $p_i(\anchor \given \z)$,
and $p_i(\z \given \anchor)$ as defined for
relation~(\ref{eq:mixture-conditional}).
Then
\begin{equation}\label{eq:mixture-z-marg}
p_i(\z) = \gauss(\z\given \vec{m}_{\z i}, \mat{V}_{\z\z i})
\end{equation}
and
\begin{equation}\label{eq:mixture-anchor-cond}
p_i(\anchor \given \z)
= \gauss(\anchor\given \vec{m}_{\anchor i}', \mat{V}_{\anchor\anchor i}')
,
\end{equation}
where
\begin{equation}\label{eq:mixture-cond-par}
\vec{m}_{\anchor i}' = \vec{m}_{\anchor i}
    + \mat{V}_{\anchor\z i} \mat{V}_{\z\z i}^{-1} (\z - \vec{m}_{\z i})
    ,
\quad
\mat{V}_{\anchor\anchor i}' = \mat{V}_{\anchor\anchor i}
    - \mat{V}_{\anchor\z i} \mat{V}_{\z\z i}^{-1} \mat{V}_{\z\anchor i}
.
\end{equation}
Here $\z$ is the observed forward data.
Substituting (\ref{eq:mixture-z-marg})-(\ref{eq:mixture-anchor-cond})
into~(\ref{eq:mixture-conditional}), we get the conditional density
\begin{equation}\label{eq:mixture-updated}
p(\anchor \given \z, \vmu, \mSigma)
\approx \sum_{i=1}^n
    v_i\,
    \gauss(\anchor\given \vec{m}_{\anchor i}',
        \mat{V}_{\anchor\anchor i}')
,\quad
\text{where }
v_i
= \frac{w_i\, p_i(\z)}{\sum_{j=1}^n w_j\, p_j(\z)}.
\end{equation}
The posterior approximation $f^{(k-1)}$ is now updated to
$f^{(k)}(\anchor) = p(\anchor \given \z)$.
This completes an iteration of the algorithm.
\end{enumerate}

Note $f^{(k)}$ is a normal mixture, just like $f^{(k-1)}$,
and is ready to be updated in the next iteration.

\subsection{Accommodating model and measurement errors}
\label{sec:accommodate-errors}

Suppose the combined model and measurement error in $\z$
is additive, as represented in~(\ref{eq:additive-error}).
Assume $\err$ is independent of $\anchor$, $\vY$,
and $\Z$.
Further assume the error is normal,
$\err \sim N(\vec{0}, \mSigma_{\err})$.
Then the
$\mat{V}_{\z\z i}$ in
(\ref{eq:mixture-z-marg}) and~(\ref{eq:mixture-cond-par}) should be replaced by
$\mat{V}_{\z\z i} + \mSigma_{\err}$.

The algorithm is able to accommodate more general errors.
For example,
non-additive errors
and
errors that might occur in certain internal steps of the forward model $\M$
can be built into the forward evaluation $\z_i = \M(\vy_i)$
of step~2 by introducing randomness according to a description
of the errors.
This way, multiple forms of errors can be accommodated.
An overall, explicit formula for the errors as a whole
is not required.
The passage from $\anchor$ to $\z$ in steps~1--2
embodies our best knowledge of the parameter-data connection:
part in the anchor parameterization
$p(\vy \given \anchor, \vmu, \mSigma)$,
part in the forward model $\M(\vy)$,
and part in errors in our implementation of $\M$ and in the measurement
of $\z$.

Going one step further,
it is possible to relax the requirement that the forward model $\M$ be ``deterministic''.
If the forward model is stochastic
\citep[see an example in][]{Besag:1991:BIR},
the parameter-data connection, as depicted
in~(\ref{eq:anchor-data-randomness}),
includes one more layer of randomness.
As far as the algorithm is concerned,
the situation is not different from inclusion of model errors.

\subsection{Comments}
\label{sec:algor-comments}

The overall idea in this algorithm may be summarized as
``simulate $p(\anchor, \Z)$, then condition on the observed $\z$.''

The algorithm does not require one to be able to draw a random sample
from the prior of $\anchor$.
Instead, a convenient initial approximation $f^{(0)}$ starts the procedure.
One only needs to be able to evaluate the prior density at any
particular value of $\anchor$.
This provides flexibilities in specifying
both the prior $\pi(\anchor)$ and the initial approximation $f^{(0)}$.

Some statistics of the distribution $f^{(k)}$
may be examined semi-analytically, taking advantage of its being a
normal mixture.
More often, one is interested in the unknown field $\vY$ or a function
thereof rather than in the parameter $\anchor$.
These may be investigated based on a sample of $\vY$
based on the posterior of $\anchor$,
as explained in
section~\ref{sec:anchor-parameterization}.

Typically,
by far the most expensive operation in this algorithm
is evaluating the forward model $\M(\vy)$.
This occurs once for each sampled value of $\anchor$.
In contrast, sampling $\anchor$ or generating field realizations
are easy,
because the distributions involved are normal or normal mixtures.

The versatility of normal mixture in approximating complex densities is
well documented
\citep{Marron:1992:EMI, West:1993:APD, McLachlan:2000:FMM}.
This approximation requires every component of
$(\anchor, \Z)$ to be a continuous variable defined on
$(-\infty, \infty)$, hence some transformations may be necessary
in the definition of $\anchor$ and $\Z$.
(See section~\ref{sec:examples} for examples.)

The dimension reduction achieved by principal component analysis
in step~3 can be significant, for example with time series data.
This facilitates the use of high-dimensional (\ie large $n_\z$)
data without worrying (too much) about the correlation between the data components.
The dimension reduction is also a stability feature.
For example, components of $\Z$ that are almost constant in the simulations
will not cause trouble.

The global structure of this algorithm bears similarities to that of
\citet{West:1993:APD}, with important differences.
While \citet{West:1993:APD} focuses on situations where the
likelihood function is known, the algorithm here completely
forgoes the requirement of a (or the) likelihood function.
This distinction has far-reaching implications for the applicability
of the proposed method.

This is an important feature of the proposed algorithm:
it does not require a likelihood function.
This feature distinguishes itself from the Markov chain Monte Carlo
(MCMC) methods.
There is a variant of MCMC
called ``approximate Bayesian computation (ABC)''
\citep[see][]{Marjoram:2003:MCM} that applies MCMC in problems without
known likelihood functions.
ABC uses a measure of the simulation-observation mismatch in
lieu of likelihood.
For multivariate data,
there are significant difficulties in how to define this measure.
The algorithm described here is very different from ABC.

In fact, this algorithm
is not intrinsically tied to the anchor parameterization or anchored
inversion.
It is a general algorithm for approximating the posterior distribution
where the likelihood is unknown but the model (which corresponds to the
likelihood) can be simulated.

\section{Automatic and adaptive choice of anchors}
\label{sec:choice-of-anchors}

The definition of anchors in~(\ref{eq:anchor-def})
is represented by the matrix $\mH$.
There are two aspects in the choice of anchors, namely
the number of anchors (\ie the number of rows in $\mH$)
and the specific definition of each anchor
(\ie the content of each row of $\mH$).
In this section we develop a procedure that
adjusts both aspects in an automatic, adaptive manner,
taking advantage of the iterative nature of the algorithm.

We partition the model domain (or numerical grid) into a number of
(equal- or unequal-sized) subsets
and define the mean value of $Y$ in each subset an anchor.
These anchors form a \emph{partitioning anchorset},
or ``anchorset'' for short.
Each anchorset is defined by a unique matrix $\mH$ as introduced before.
This definition is instrumental in the adaptive procedure,
to be presented shortly,
which examines a number of anchorsets and chooses the one
that has the best predicted performance in the upcoming iteration.

\subsection{A measure of model performance}

The choice of anchorset is a model selection problem,
because any anchorset represents a particular
parameterization, or ``model'', for the field $\Y(\x)$.
To facilitate comparison of models, we measure the performance of a
model by a kind of integrated predictive likelihood defined as
\[
L_I = \int_{\Anchor}
    p(\z \given \anchor, \vmu, \mSigma)\,
    f^{(k)}(\anchor) \diff \anchor,
\]
where $f^{(k)}(\anchor)$ is the approximate posterior density of
$\anchor$ obtained in the $k$th iteration.
This measure is analogous to Bayes factors
\citep{Kass:1995:BF, George:2006:BMS} or, in particular,
``posterior'' Bayes factors
\citep{Aitkin:1992:PBF}.
However, the current setting differs from common (Bayesian) model
selection in several ways.
For example,
the likelihood
$p(\z \given \anchor, \vmu, \mSigma)$ is unknown, and is being estimated;
the ultimate subject is not really the model parameter $\anchor$,
but rather the field $\Y$;
the algorithm is iterative, which makes use of the data $\z$ in each
iteration and corrects for the repeated use of data by importance
weighting.

As an approximation,
we assume the predictive distribution,
$\int p(\z \given \anchor, \vmu, \mSigma)\, f^{(k)}(\anchor)\diff \anchor$,
is normal.
The predictions
$\z_1,\dotsc,\z_n$ generated in step~2 of the algorithm is a
random sample of this distribution.
Based on the sample mean $\overline{\z}$ and sample covariance
matrix $\mat{S}_z$ of $\z_1,\dotsc,\z_n$,
an approximation of $\log L_I$ could be computed by
\[
L = \log \gauss\bigl(\z\given \overline{z}, \mat{S}_z\bigr)
,
\]
where $\z$ is the observation.

The measure $L$ is subject to a fair amount of
uncertainty arising from several sources.
First, it is computed based on \emph{random samples}
$\anchor_1,\dotsc,\anchor_n$ and $\z_1,\dotsc,\z_n$.
Second, each $\z$ is a random draw conditional on its corresponding
$\anchor$ (via a random draw of $\vy$).
Third,
the normal assumption for $\Z$ may not be fully justifiable.
Fourth, the sample size ($n$) is rather moderate in view of the
dimensions of the random variables ($n_\anchor$ and $n_\z$).
In experiments we used $n$ in the thousands and $n_\anchor + n_\z$
in the tens or hundreds.
In fact, when $n_\anchor + n_\z$ is relatively large (say about 100),
it is not rare that the empirical covariance $\mat{S}_z$
is not invertible.
In view of these subtleties that harm the robustness of $L$,
we further simplify and take
\begin{equation}\label{eq:log-int-likely}
L^* = \sum_{j=1}^{n_z} \log \gauss
    \bigl(\z[j]\biggiven \overline{z[j]}, s^2_{z[j]}\bigr)
,
\end{equation}
where $[j]$ indicates the $j$th dimension of $\Z$.
This approximation ignores correlations between the components of $\Z$.

\subsection{Predicting the performance of alternative anchorsets}

Denote the anchorset at the beginning of the $k$th iteration
by $\mH$.
Towards the end of this iteration (somewhere in step~5),
we consider whether to switch to an alternative anchorset,
say $\mH_*$,
in the next iteration.
The decision to switch or not will be made after comparing the
\emph{predicted performances} of the model in the next iteration
using anchors $\mH$ and $\mH_*$, respectively.
Let the two performances be denoted by
$L_{k+1}^*(\mH)$ and
$L_{k+1}^*(\mH_*)$,
respectively.

The model~(\ref{eq:field-prior}),
the prior~(\ref{eq:anchor-prior}),
and
the relation~(\ref{eq:Y-given-anchor})
suggest that $\anchor_*$ conditional on $\anchor$ has a normal distribution,
which we denote by $\pi(\anchor_* \given \anchor, \vmu, \mSigma)$
to stress the fact that it is derived from the ``prior'' relations.
Analogously,
we write $\pi(\anchor \given \anchor_*, \vmu, \mSigma)$ and
$\pi(\anchor, \anchor_* \given \vmu, \mSigma)$.

The distribution $f^{(k-1)}(\anchor)$ induces a distribution
for $\anchor_*$, which we denote by $f_*^{(k-1)}(\anchor_*)$.
In fact,
the sampling in steps~1--2 of the $k$th iteration
also generates a sample of the alternative anchors,
$\{\anchor_{*i} = \mH_*\vy_i\}_{i=1}^n$,
and the sample is a random sample of $f_*^{(k-1)}(\anchor_*)$
by the definition of $f_*^{(k-1)}(\anchor_*)$.
Alternatively,
we can consider the sample $\{\anchor_{*i}\}$
to be obtained via
$f_*^{(k-1)}(\anchor_*)
= \int_{\Anchor} \pi(\anchor_* \given \anchor, \vmu, \mSigma)\,
    f^{(k-1)}(\anchor) \diff \anchor$.

Suppose we continue to use $\mH$ in the $(k+1)$th iteration.
In step~5 of the $k$th iteration,
$f^{(k-1)}(\anchor)$ is updated to $f^{(k)}(\anchor)$.
The distribution $f^{(k)}(\anchor)$ will be used in steps~1--2
of the $(k+1)$th iteration to generate a random sample of $\Z$,
which is then used to calculate the performance indicator
$L_{k+1}^*(\mH)$.
Due to the high computational cost of the forward model,
however, we do not want to actually generate this sample of $\Z$
before committing to using the anchors $\mH$ in the $(k+1)$th iteration.
The idea then is to use the sample
$\z_1,\dotsc,\z_n$ created in step~2
of the $k$th iteration, which is based on $f^{(k-1)}(\anchor)$,
as an importance sample from the distribution of $\Z$
that would be generated in step~2 of the $(k+1)$th iteration,
which would be based on $f^{(k)}(\anchor)$.
The importance weights are proportional to
$f^{(k)}(\anchor_i) / f^{(k-1)}(\anchor_i)$, $i=1,\dotsc,n$.
This weighted sample of $\Z$ then provides
weighted sample means $\overline{\z[j]}$ and
weighted sample variances $s^2_{\z[j]}$
to be used in~(\ref{eq:log-int-likely}) for predicting
$L_{k+1}^*(\mH)$.

Now suppose we switch to $\mH_*$ in the $(k+1)$th iteration.
First,
at the end of step~5 of the $k$th iteration,
we need to obtain $f_*^{(k)}(\anchor_*)$ instead of $f^{(k)}(\anchor)$.
Second,
the distribution $f_*^{(k)}(\anchor_*)$ would give rise to
a sample of $\Z$ in step~2 of the $(k+1)$th iteration,
but we again would like to use the sample
$\z_1,\dotsc,\z_n$ created in step~2 of the $k$th iteration
as a \emph{weighted} replacement.
We address these two steps now.

\begin{enumerate}
\item[(1)]
We would like to view the sample $\{\anchor_{*i}\}$
as an importance sample from
$\pi(\anchor_*)$,
then we would be able to conduct steps~4--5 for
$\anchor_*$ (instead of $\anchor$) and obtain
$f_*^{(k)}(\anchor_*)$.
The question is: What are the importance weights?

If the anchorsets $\mH$ and $\mH_*$ have no common anchors,
then
\begin{equation}\label{eq:adapt-weight-1}
\pi(\anchor_*)
= \int_{\Anchor}
    \pi(\anchor_* \given \anchor, \vmu, \mSigma)\,
    \pi(\anchor)
    \diff \anchor
= \int_{\Anchor}
    \pi(\anchor_* \given \anchor, \vmu, \mSigma)\,
    \frac{\pi(\anchor)}{f^{(k-1)}(\anchor)}
    f^{(k-1)}(\anchor)
    \diff \anchor
.
\end{equation}
This suggests that the importance weights are proportional to
$\frac{\pi(\anchor)}{f^{(k-1)}(\anchor)}$.

If $\mH$ and $\mH_*$ share some common anchors, then write
$\anchor = (\anchor_\square, \anchor_\diamond)$ and
$\anchor_* = (\anchor_\square, \anchor_\bigstar)$.
Let
$f^{(k-1)}(\anchor_\square)$ and
$f^{(k-1)}(\anchor_\diamond \given \anchor_\square)$ denote
the marginal and conditional distributions induced
by $f^{(k-1)}(\anchor)$.
We have
\begin{equation}\label{eq:adapt-weight-2}
\begin{split}
\pi(\anchor_*)
&= \pi(\anchor_\square)\, \pi(\anchor_\bigstar \given \anchor_\square, \vmu, \mSigma)
\\
&= f^{(k-1)}(\anchor_\square)
    \frac{\pi(\anchor_\square)}{f^{(k-1)}(\anchor_\square)}
  \int_{\Anchor_\diamond}
    \pi(\anchor_\bigstar \given \anchor_\diamond, \anchor_\square, \vmu, \mSigma)\,
    \frac{\pi(\anchor_\diamond \given \anchor_\square, \vmu, \mSigma)}
        {f^{(k-1)}(\anchor_\diamond \given \anchor_\square)}
    f^{(k-1)}(\anchor_\diamond \given \anchor_\square)
    \diff \anchor_\diamond
.
\end{split}
\end{equation}
Noticing that the sample $\{\anchor_{*i}\}$ is obtained via
$f^{(k-1)}(\anchor_\square)\,
 f^{(k-1)}(\anchor_\diamond \given \anchor_\square)\,
 \pi(\anchor_\bigstar \given \anchor_\diamond, \anchor_\square, \vmu, \mSigma)$,
the above suggests the importance weights are proportional to
$\frac{\pi(\anchor_\square)}{f^{(k-1)}(\anchor_\square)}
 \frac{\pi(\anchor_\diamond \given \anchor_\square, \vmu, \mSigma)}
    {f^{(k-1)}(\anchor_\diamond \given \anchor_\square)}
= \frac{\pi(\anchor)}{f^{(k-1)}(\anchor)}$.

Relations (\ref{eq:adapt-weight-1})--(\ref{eq:adapt-weight-2})
show that the weights
are proportional to $\pi(\anchor_i) / f^{(k-1}(\anchor_i)$
when $\{\anchor_{*i}\}$ are taken as an importance sample from $\pi(\anchor_*)$.
These are the same weights (\ie the $w_i$s) had we continued to use
$\anchor$ in steps~4--5.
In other words,
the weights do not depend on the specific choice of $\mH_*$.
Using these weights and the sample $\{\anchor_{*i}, \z_i\}$,
steps~4--5 of the $k$th iteration produce
$f_*^{(k)}(\anchor_*)$.

\item[(2)]
After deriving $f_*^{(k)}(\anchor_*)$,
we would view $\{\z_i\}$, obtained in step~2 of the $k$th iteration,
as an importance sample
of the $\Z$ that would be generated starting with $f_*^{(k)}(\anchor_*)$
in steps~1--2 of the $(k+1)$th iteration.
The importance weights are equal to those
of $\{\anchor_{*i}\}$ obtained in the $k$th iteration
as an importance sample from $f_*^{(k)}(\anchor_*)$.

If the anchorsets $\mH$ and $\mH_*$ have no common anchors,
then
\begin{equation}\label{eq:adapt-weight-3}
\begin{split}
f_*^{(k)}(\anchor_*)
&= \frac{f_*^{(k)}(\anchor_*)}{\pi(\anchor_*)}
  \int_{\Anchor}
        \pi(\anchor_* \given \anchor, \vmu, \mSigma)
        \,\pi(\anchor)
        \diff \anchor
\\
&= \frac{f_*^{(k)}(\anchor_*)}{\pi(\anchor_*)}
  \int_{\Anchor}
        \pi(\anchor_* \given \anchor, \vmu, \mSigma)
        \frac{\pi(\anchor)}{f^{(k-1)}(\anchor)}
        f^{(k-1)}(\anchor)
        \diff \anchor
.
\end{split}
\end{equation}
This suggests the importance weights are proportional to
\begin{equation}\label{eq:anchorset-weights}
\frac{f_*^{(k)}(\anchor_*) / \pi(\anchor_*)}
    {f^{(k-1)}(\anchor) / \pi(\anchor)}
.
\end{equation}

If $\mH$ and $\mH_*$ share some common anchors, then,
using the notation introduced above, we have
\begin{equation}\label{eq:adapt-weight-4}
\begin{split}
f_*^{(k)}(\anchor_*)
&= f_*^{(k)}(\anchor_\square)\,
    f_*^{(k)}(\anchor_\bigstar \given \anchor_\square)
\\
&= f^{(k-1)}(\anchor_\square)
    \frac{f_*^{(k)}(\anchor_\square)}
        {f^{(k-1)}(\anchor_\square)}
    \frac{f_*^{(k)}(\anchor_\bigstar \given \anchor_\square)}
        {\pi(\anchor_\bigstar\given \anchor_\square)}
    \int_{\Anchor_\diamond}
    \pi(\anchor_\bigstar \given \anchor_\diamond,
        \anchor_\square, \vmu, \mSigma)\,
    \pi(\anchor_\diamond \given \anchor_\square)
    \diff \anchor_\diamond
\\
&= f^{(k-1)}(\anchor_\square)
    \frac{f_*^{(k)}(\anchor_\square)}
        {f^{(k-1)}(\anchor_\square)}
    \frac{f_*^{(k)}(\anchor_\bigstar \given \anchor_\square)}
        {\pi(\anchor_\bigstar\given \anchor_\square)}
\\
&\phantom{\mathrel{=} }\;\;
    \int_{\Anchor_\diamond}
    \pi(\anchor_\bigstar \given \anchor_\diamond,
        \anchor_\square, \vmu, \mSigma)\,
    \frac{\pi(\anchor_\diamond \given \anchor_\square)}
        {f^{(k-1)}(\anchor_\diamond \given \anchor_\square)}
    f^{(k-1)}(\anchor_\diamond \given \anchor_\square)
    \diff \anchor_\diamond
,
\end{split}
\end{equation}
suggesting the importance weights are proportional to
\begin{equation}\label{eq:anchorsets-weights-alt}
\frac{f_*^{(k)}(\anchor_\square)}
    {f^{(k-1)}(\anchor_\square)}
\frac{f_*^{(k)}(\anchor_\bigstar \given \anchor_\square)}
    {\pi(\anchor_\bigstar\given \anchor_\square)}
\frac{\pi(\anchor_\diamond \given \anchor_\square)}
    {f^{(k-1)}(\anchor_\diamond \given \anchor_\square)}
= \frac{f_*^{(k)}(\anchor_*) /
            \pi(\anchor_\bigstar \given \anchor_\square)}
        {f^{(k-1)}(\anchor) /
            \pi(\anchor_\diamond \given \anchor_\square)}
= \frac{f_*^{(k)}(\anchor_*) / \pi(\anchor_*)}
        {f^{(k-1)}(\anchor) / \pi(\anchor)}
.
\end{equation}

The weights turn $\{z_i\}$ into a weighted sample
in computing $\overline{z[j]}$ and $s^2_{z[j]}$
for predicting $L^*(\mH_*)$ by~(\ref{eq:log-int-likely}).
\end{enumerate}

Next,
if it turns out that
$L^*(\mH_*) > L^*(\mH)$, we take $f_*^{(k)}(\anchor_*)$ as the output of
the $k$th iteration, with the anchorset $\mH_*$.
Otherwise, the output is $f^{(k)}(\anchor)$ with the anchorset $\mH$.

A remaining question is,
how do we identify,
in a way that is flexible and efficient,
candidate alternative anchorsets (\ie $\mH_*$'s) to examine?

\subsection{A strategy for defining alternative anchorsets}

Suppose the current anchorset,
represented by $\mH$, contains $n_{\anchor}$ anchors.
Recall the definition of anchorset:
the model domain is partitioned into $n_{\anchor}$ sub-regions;
the mean value of $Y$ in each sub-region is designated an anchor.
Let us call each sub-region the ``support'' of the corresponding anchor.

We consider $n_{\anchor}$ alternative anchorsets,
denoted by $\mH_1,\dotsc,\mH_{n_{\anchor}}$,
each containing $n_{\anchor} + 1$ anchors.
The anchorset $\mH_{j}$, $j=1,\dotsc,n_\anchor$,
is identical to the original anchorset $\mH$ except that
the support of the $j$th anchor in $\mH$ is split into two sub-regions,
defining two new anchors.

The $2 n_{\anchor}$ newly-created ``small'' anchors
resulting from splitting each anchor in $\mH$
form an ``umbrella'' anchorset, denoted by $\mH_*$,
which defines anchor vector $\anchor_* = \mH_*\vY$.
Similarly defined are anchor vectors
$\anchor = \mH\vY$, $\anchor_1 = \mH_1\vY$,...,
$\anchor_{n_\anchor} = \mH_{n_\anchor}\vY$.
It can be seen that each of
$\anchor$, $\anchor_1$,..., $\anchor_{n_\anchor}$
is a linear function of $\anchor_*$.
In steps~3--5 of the algorithm,
we estimate the density of $\anchor_*$.
(To estimate $f^{(k)}(\anchor_*)$,
extract the sample $\{\mH_*\vy_i\}$, then use
the weights $\{w_i\}$ in steps~3--5.)
Because the estimate is a normal mixture,
the densities of
$\anchor$, $\anchor_1$,..., $\anchor_{n_\anchor}$
follow immediately.
Based on these densities,
we predict the $L^*$ for each of these $n_\anchor + 1$ anchorsets
and choose the anchorset that achieves the largest $L^*$.
Steps~3--5 of the algorithm are then repeated to estimate $f^{(k)}$
for the chosen anchorset.

By this strategy,
steps 3--5 of the algorithm are performed twice
in order to compare and choose from $n_{\anchor} + 1$ anchorsets,
including the original and the alternatives.
As iterations proceed,
more alternative anchorsets are considered in each iteration
as the size of (\ie number of anchors contained in)
the currently used anchorset grows.
The computational cost, however, barely increases.
This efficient strategy takes advantage of the fact that the weights
$\{w_i\}$ remain the same for all anchorsets.

An obvious extension to this strategy allows an alternative anchorset
to split more than one anchor (say up to two) in the original anchorset.
This extension examines more alternative anchorsets without
substantially increasing the computational cost,
because the umbrella anchorset $\mH_*$ remains unchanged.

As this adaptive procedure suggests,
the inversion exercise begins with a low number of anchors
and gradually increases the number of anchors in iterations.
In particular,
we may begin with an anchorset that bisects
each dimension of the space;
that is, use 2, 4, and 8 anchors in 1D, 2D, and 3D spaces,
respectively.
This completely relieves the user
from the burden of specifying an initial anchorset.
More anchors are introduced in the iterations
as more computation has been invested in the task.
New anchors tend to be introduced in regions of the model domain
where a more detailed representation of the $Y$ field
is predicted to bring about most improvement
to the model's capability in reproducing the observation $\z$.
The partition of the model domain by the anchorset
is not uniform---it is tailored to the unknown field $Y$,
the forward process $\M$, and existing anchors in an
automatic, adaptive, and evolving fashion.

\subsection{Modifications to the algorithm in section~\ref{sec:algorithm}}

In steps 3--5,
the original $\anchor$ is replaced by the umbrella $\anchor_*$
(obtained from the sample $\{\vy_i\}$ via $\mH_*\vy_i$),
which is usually twice as long as $\anchor$.
At the end of step~5,
$p(\anchor_*\given \z, \vmu, \mSigma)$ is obtained.

Step~5 now does not conclude an iteration.
Display~(\ref{eq:mixture-updated}) is followed by the following new
step.

\begin{enumerate}
\item[6.] \emph{Selecting an anchorset and deriving its distribution.}

From the estimated density $p(\anchor_* \given \z, \vmu, \mSigma)$,
given by~(\ref{eq:mixture-updated}),
derive the density of each alternative anchorset
as well as the original anchorset,
then predict $L^*$ for each of these anchorsets.

Pick the anchorset that achieves the largest $L^*$.
Use this anchorset to repeat steps~3--5 and derive its
density $p(\anchor \given \z, \vmu, \mSigma)$.
This density becomes $f^{(k)}(\anchor)$,
where $\anchor$ corresponds to the picked anchorset.

This concludes an iteration of the algorithm.
\end{enumerate}

\section{Extensions}
\label{sec:extensions}

\subsection{Geostatistical parameterization of the field}
\label{sec:geostat-extension}

The mean vector $\vmu$ and
covariance matrix $\mSigma$ of the field
may be parameterized in
conventional geostatistical forms.
Because of the existence of anchors,
sophisticated trend models, such as polynomials in spatial coordinates,
are usually unnecessary.
One may model the mean $\vmu$ by a global constant:
\[
\vmu = \beta
.
\]
The stationary covariance may be parameterized as
\[
\cov\bigl(\Y(\x_1), \Y(\x_2)\bigr)
= \eta^2 \rho(\x_1 - \x_2; \varphi)
,
\]
where $\eta^2$ is variance
and $\varphi$ contains parameters pertaining to
range, smoothness, nugget, geometric anisotropy, and so on.
We leave specifics of this parameterization open.
See section~\ref{sec:geostat-par} for an example of a
particular form of this parameterization.

A requirement imposed by the proposed inversion method is that
all unknown parameters are defined on supports that can be mapped onto
$(-\infty, \infty)$.
This usually means that the support of each parameter should
be in one of the following forms:
$(-\infty,\infty)$, $(-\infty, c)$, $(c, \infty)$,
and $(c_1, c_2)$.
Once this requirement is satisfied,
all parameters are transformed to be defined on
$(-\infty, \infty)$, hence the posterior distribution
can be approximated by a normal mixture.
Note the support of each anchor is already $(-\infty, \infty)$
because the support of $\Y$ is $(-\infty, \infty)$.

Let us denote the geostatistical parameters
$\beta$, $\eta^2$, and $\varphi$, transformed onto $(-\infty, \infty)$,
by $\anchor_1$
and re-label the anchor parameters $\anchor_2$.
The complete parameter vector is now $\anchor = (\anchor_1, \anchor_2)$.

The parameterization of $\Y$ described in
section~\ref{sec:anchor-parameterization},
or (\ref{eq:Y-given-anchor}) in particular,
now is written as
$p(\vy \given \anchor)$, \ie
$p(\vy \given \anchor_1, \anchor_2)$,
in which $\anchor_1$ determines $\vmu$ and $\mSigma$,
and $\anchor_2$ plays the role of the $\anchor$ in~(\ref{eq:Y-given-anchor}).
The expanded parameter vector
retains the simplicity of a geostatistical description of the random field.
In the meantime it enables,
via~(\ref{eq:Y-given-anchor}),
sophisticated mean and covariance structures for $\Y$
beyond the expressing capability of the geostatistical parameters alone.
To some extent,
one may say that the geostatistical parameters capture ``global'' features,
whereas the anchor parameters capture ``local'' features.

A prior for $\anchor$ can be specified by
\begin{equation}\label{eq:anchor-geostat-prior}
\pi(\anchor)
= \pi(\anchor_1)\, \pi(\anchor_2\given \anchor_1)
,
\end{equation}
where $\pi(\anchor_1)$ is a prior for the geostatistical parameters,
and
$\pi(\anchor_2\given \anchor_1)
= \gauss(\anchor_2 \given \mH\vmu, \mH\mSigma\transpose{\mH})$
in which $\vmu$ and $\mSigma$ are determined by $\anchor_1$.
The posterior is now written as
$p(\anchor \given \z)$ instead of $p(\anchor \given \z, \vmu, \mSigma)$.

In the algorithm in section~\ref{sec:inference},
all the $\anchor$'s
should be understood as the expanded parameter vector,
$\anchor = (\anchor_1, \anchor_2)$.
The overall idea of the algorithm can be summarized as
``simulate $p(\anchor_1, \anchor_2, \Z)$,
then condition on the observed $\z$.''

In step~1 of the algorithm,
the prior $\pi(\theta_i)$ is now given by~(\ref{eq:anchor-geostat-prior}).
In step~2,
$p(\vy \given \anchor_i, \vmu, \mSigma)$
is now $p(\vy \given \anchor_i)$,
in which
$\anchor_i$ contains both anchor parameters and
geostatistical parameters,
the latter providing
$\vmu$ and $\mSigma$.

Similarly in
section~\ref{sec:choice-of-anchors},
the $\anchor$ and $\anchor_*$ should be understood as
the extended parameter vectors,
and their priors are given by~(\ref{eq:anchor-geostat-prior}).
In the meantime $\vmu$ and $\mSigma$ should be dropped because they are now
determined by $\anchor_1$.
Note that the $\anchor_1$ part is not affected by the search for
alternative $\anchor_2$, hence
$\anchor_1$ plays the role of the $\anchor_\square$ in
(\ref{eq:adapt-weight-2}), (\ref{eq:adapt-weight-4}),
and~(\ref{eq:anchorsets-weights-alt}).

\subsection{Using linear data}
\label{sec:linear-data}

Besides the nonlinear data $\z$,
one may have some direct measurements of the spatial attribute $\Y$
itself or, more generally, measurements of some linear function of $\Y$.
In the groundwater example in section~\ref{sec:intro},
linear data may include
direct measurements of local-scale hydraulic conductivity
and covariates such as grain-size distribution and
core-support geophysical properties,
which provide estimates of local hydraulic conductivity via empirical
relations.
In the geophysical example in section~\ref{sec:intro},
linear data may be mechanic properties of the elastic medium
at the ground surface.
In the atmospheric example in section~\ref{sec:intro},
linear data may be direct monitoring of CO$_2$ sources on the ground
surface or covariates that provide estimates of CO$_2$ sources.

Linear data enter the proposed inversion procedure at two places:
(1) the prior of anchors, $\pi(\anchor)$, not only relies on
the mean $\vmu$ and covariance $\mSigma$,
but is also conditioned on the linear data;
(2) while simulating field realizations conditional on values of the
anchor parameters, the simulation is in addition conditioned on the
linear data.
The modified prior can be alternatively viewed as
the prior $\pi(\anchor)$, ignoring the linear data,
multiplied by the likelihood of $\anchor$
with respect to the linear data.

Let the linear data be denoted by
$\ell = \mat{L}\vy$.
(Naturally, $\mat{L}$ shares no common row with $\mH$,
the definition of anchors.)
If the geostatistical extension described in
section~\ref{sec:geostat-extension} is not used,
the prior of $\anchor$ given by~(\ref{eq:anchor-prior})
is replaced by
\begin{equation}\label{eq:anchor-linear-prior}
\pi(\anchor)
\propto
p(\anchor \given \vmu, \mSigma)\,
    p(\ell \given \anchor, \vmu, \mSigma)
= p(\anchor, \ell \given \vmu, \mSigma)
= \gauss\biggl(
    \begin{pmat}[{}] \anchor \\ \cr\- \ell \cr\end{pmat}
    \;
    \biggm|
    \;
    \begin{pmat}[{}] \mH \\ \cr\- \mat{L} \cr\end{pmat} \vmu
    ,\,
    \begin{pmat}[{|}]
        \mH \mSigma \transpose{\mH} & \mH\mSigma\transpose{\mat{L}}
        \\ \cr\-
        \mat{L} \mSigma \transpose{\mH} & \mat{L}\mSigma\transpose{\mat{L}}
        \cr\end{pmat}
    \biggr)
.
\end{equation}
The conditional distribution of the field $\vY$ given
the anchor values is now
$
p(\vy \given \anchor, \ell, \vmu, \mSigma)
$.
This can be handled by combining $\anchor$ and $\ell$
into a single set of linear condition
and using a formula analogous to~(\ref{eq:Y-given-anchor}).

In this situation,
the target of the exercise is the posterior
$p(\anchor \given \z, \ell, \vmu, \mSigma)$.
The overall idea of the algorithm can be summarized as
``simulate $p(\anchor, \Z \given \ell, \vmu, \mSigma)$,
then condition on the observed $\z$.''

If the geostatistical extension is in place, let us denote
$\anchor = (\anchor_1, \anchor_2)$ as in
section~\ref{sec:geostat-extension}.
The prior in~(\ref{eq:anchor-geostat-prior}) is replaced by
\begin{equation}\label{eq:anchor-geostat-linear-prior}
\pi(\anchor)
\propto
\pi(\anchor_1)\,
    p(\anchor_2 \given \anchor_1)\,
    p(\ell \given \anchor_1, \anchor_2)
= \pi(\anchor_1)\, p(\anchor_2, \ell \given \anchor_1)
.
\end{equation}
The conditional $p(\anchor_2, \ell \given \anchor_1)$
is normal, analogous to~(\ref{eq:anchor-linear-prior})
except that $\vmu$ and $\mSigma$ are determined by $\anchor_1$.
The conditional distribution of $\vY$
given the model parameter $\anchor$ is conditioned on $\ell$ in addition
to $\anchor$,
that is,
$p(\vy \given \anchor, \ell)
= p(\vy \given \anchor_2, \anchor_1, \ell)$,
where $\anchor_2$ and $\ell$ are linear conditions,
and $\anchor_1$ provides $\vmu$ and $\mSigma$.
This is again similar to~(\ref{eq:Y-given-anchor}).

In this situation,
the target of the exercise is the posterior
$p(\anchor_1, \anchor_2 \given \z, \ell)$.
The overall idea of the algorithm can be summarized as
``simulate $p(\anchor_1, \anchor_2, \Z \given \ell)$,
then condition on the observed $\z$.''

\section{Examples}
\label{sec:examples}

We will present three examples of scientific applications using synthetic data.
The first two examples are in one-dimensional space,
whereas the third is in two-dimensional space.
In all three examples,
the unknown physical property of interest is positive by definition,
hence we take $\Y$ to be the logarithm of the physical property,
and treat $\Y$ as a Gaussian process.
The examples demonstrate a number of aspects
of the anchored inversion methodology.
In example~1,
the forward data are measurements distributed in space.
In example~2,
the forward data are time-series measurements at a boundary point of the
model domain.
In example~3,
the forward data are neither time series nor attached to
specific spatial locations.
In fact, all components of the forward-data in the examples
are functions of the entire field,
even if their measurement is associated with specific locations or times.
Example~1 also uses a linear datum (\ie, a direct measurement of $\Y$
at one location).
In example~2,
the forward function is a composition of two separate field processes,
giving two datasets
(to be combined to form the forward data vector)
that are of incomparable physical natures.
In example~3,
the inversion algorithm is executed in a numerical grid that is coarser
than the synthetic true field, introducing systematic model error.
In all examples,
there exists substantial correlation between
components of the forward data (\ie variable $Z$).

In all the examples,
the data were assumed to be error-free.
(In example~3, we knew there was model error but chose to ignore it.)
Because the true field is known in these synthetic examples,
field simulations compared with the true field demonstrate
the performance of the inversion.
Another assessment comes from comparing
predictions of the forward data in simulated fields
with the observed forward data.

In all examples,
the algorithm was run for 20 iterations with sample sizes (the $n$s)
2400, 1950, 1612,..., 610, 608, totaling 19176 samples
(which is also the total number of forward model runs).
The initial anchors are designated by evenly bisecting each spatial
dimension (hence the algorithm starts with 2 and 4 anchors in
1-D and 2-D examples, respectively).
As iteration proceeds,
anchor designations evolve automatically
as described in section~\ref{sec:choice-of-anchors}.
In each iteration we considered splitting one of the existing anchors,
hence the number of anchors increased by 0 or 1 in any single iteration.

Before delving into individual examples,
in sections~\ref{sec:geostat-par}--\ref{sec:diagnostics}
we address some general preparations that are used in all examples.
In these two sections,
both $\Y$ and $\Z$ are generic symbols for the spatial variable and
forward data, respectively, as
the actual meanings of both depend on the specific example.

\subsection{Geostatistical parameterization, prior specification,
    and initial approximation}
\label{sec:geostat-par}

For all the examples,
we used a conventional geostatistical formulation
for the field mean $\vmu$ and covariance $\mSigma$.
Following the notation in section~\ref{sec:geostat-extension},
the geostatistical parameters, anchor parameters, and the complete
parameter vector are denoted by $\anchor_1$, $\anchor_2$,
and $\anchor = (\anchor_1,\anchor_2)$, respectively.
The same procedures were followed in all examples to
specify the prior for $\anchor_1$
and construct the initial approximation $f^{(0)}$.
Recall that the methodology is open with respect to these elements.
However,
the specific choices made for these examples are sensible
starting points in general applications.

The parameterization of the field $\Y$ (before conditioning on anchors)
includes a global mean $\beta$ and a covariance function
\begin{equation}
\begin{split}
\cov\bigl(Y(x_1), Y(x_2)\bigr)
&= (1 - \tau)\eta^2 \rho(|x_1 - x_2|) + \tau \eta^2 I(x_1 = x_2)
\\
&= (1 - \tau)\eta^2 \biggl(1 + \frac{|x_1 - x_2|}{\lambda}\biggr)
        \exp\biggl(-\frac{|x_1 - x_2|}{\lambda}\biggr)
    + \tau \eta^2 I(x_1 = x_2)
,
\end{split}
\end{equation}
in which $\eta^2$ is variance, $\tau \in [0,1)$ is ``nugget'',
$I(\cdot)$ is the indicator function assuming the value 1
if its arguments is true and 0 otherwise,
$\lambda$ is ``range'' or ``scale'',
and $\rho$ is an isotropic correlation function.
The particular form of $\rho$ adopted above
is a special case of the Mat\'ern correlation function
with fixed smoothness $\kappa=1.5$.
This correlation function has been used by
\citet{Zhang:2007:ACM} in modeling hydrodynamics.
In-depth discussion on the Mat{\'ern} family of correlation functions can
be found in \citet{Stein:1999:ISD}.
In sum, the geostatistical formulation uses a parameter vector
$(\beta, \lambda, \eta^2, \tau)
\in (-\infty,\infty) \times (0,\infty)
    \times (0, \infty) \times [0,1)$,
in the natural unit of each component.
(Geometric anisotropy and the option of taking $\kappa$
as an unknown parameter are also implemented.)
The following prior is adopted for these parameters
(see \citet{Zhang:2012:IPI} for details):
\[
\pi(\beta, \lambda, \eta^2, \tau)
= \operatorname{gamma}(\lambda; \cdot) \,
    (\eta^2)^{-1}\,
    \operatorname{beta}(\tau; \cdot)
.
\]

To use the proposed algorithm,
each parameter component is transformed onto
$(-\infty,\infty)$, leading to the geostatistical parameter vector
\[
\anchor_1
= \biggl(
    \beta,\,
    \log \lambda,\,
    \log\eta^2,\,
    \log\Bigl(\frac{\tau}{1-\tau}\Bigr)
    \biggr)
.
\]
The prior of $\anchor_1$
is determined by the prior above (which is in terms of the parameters in
their ``natural'' units) and the transformations.
This prior is used in
(\ref{eq:anchor-geostat-prior})
or~(\ref{eq:anchor-geostat-linear-prior})
(depending on the presence or absence of linear data)
to specify a prior for the full model parameter $\anchor$.
Note that a prior for the anchor parameters, $\anchor_2$,
is determined by (\ref{eq:anchor-geostat-prior})
or~(\ref{eq:anchor-geostat-linear-prior}); no user intervention is
required.

The initial approximation $f^{(0)}(\anchor)$ was taken
to be a multivariate normal distribution that is fairly diffuse.

\subsection{Diagnostics}
\label{sec:diagnostics}

Roughly speaking,
the integrated likelihood $L^*$
defined by~(\ref{eq:log-int-likely})
measures how well the model reproduces,
or predicts, the observations.
Here, ``reproduction'' refers to the output of the
forward model evaluated at $Y$ fields simulated
according to the posterior distribution of $\anchor$.
In real-world applications,
a comparison of such model predictions with the actual observations
provides one of the more concrete assessments of the model.
An increasing $L^*$ in iterations suggests the model is improving.
Hence $L^*$ can be monitored as a diagnostic.

Another summary of this prediction-observation comparison
is as follows.
Consider the sample
$\{z_1,\dotsc,z_n\}$ created in step~2 of the very first iteration
(that is, based on $f^{(0)}(\anchor)$).
For dimension $j$ of $Z$,
where $j=1,\dotsc,n_\z$,
compute the median absolute difference
between the sample and the observation,
and denote it by
$\text{\tt mad}^{(0)}_{[j]}$.
Later in the $i$th iteration,
compute $\text{\tt mad}^{(i)}_{[j]}$
and obtain the ratio
$r_{\text{\tt mad}}^{(i)[j]}
= \text{\tt mad}^{(i)}_{[j]} / \text{\tt mad}^{(0)}_{[j]}$.
This ``mad ratio'' for each dimension of $Z$
is expected to decrease in iterations until it reaches a stable level.
In each iteration,
summaries such as the median and the maximum of the $n_\z$
mad ratios are reported as diagnostics.

As a diagnostic,
the ``mad ratio'' is bounded below by 0
and rarely exceeds 1.
This makes it easier to interpret than $L^*$.

\subsection{Example 1: groundwater flow}
\label{sec:example-1}

\begin{figure}
\includegraphics[scale=.8]{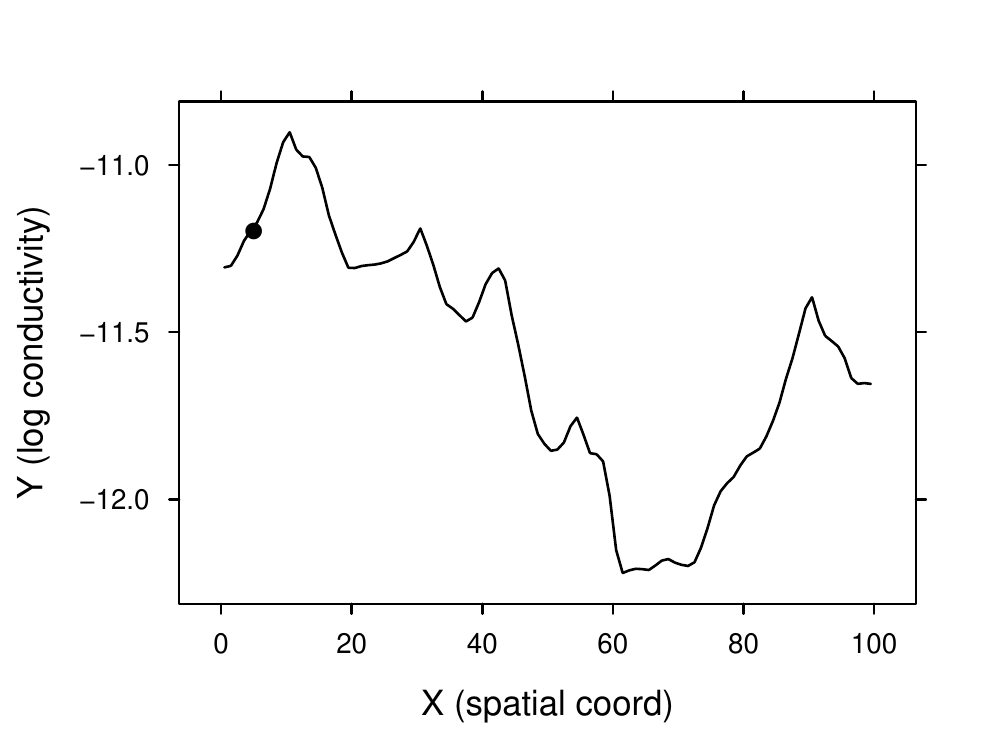}
\caption{Synthetic field of log hydraulic conductivity (meter/second)
    for example~1 in section~\ref{sec:examples}.
    The dot on the curve indicates the single direct measurement
    (acting as available linear data).}
\label{fig:darcy-field}
\end{figure}

Consider the groundwater flow example mentioned in
section~\ref{sec:intro} in one-dimensional space.
Denote hydraulic conductivity by $K$ (\unit{m s^{-1}})
and hydraulic head by $h$ (m).
We took $\Y = \log K$ (because $K$ is positive)
and modeled $\Y$ as a one-dimensional Gaussian
process.
The synthetic log-conductivity field is shown in
figure~\ref{fig:darcy-field}.
The field is composed of 100 grid points;
this constitutes the model domain.

The one-dimensional, steady-state groundwater flow in saturated zone
is described by the following differential equation
\cite[chap.~5]{Schwartz:2003:FGW}:
\begin{equation}\label{eq:ex1-forward}
\frac{\mathrm{d}}{\mathrm{d}x} \left(
    K \frac{\mathrm{d}h}{\mathrm{d}x} \right) = 0
,
\end{equation}
where $x$ is spatial coordinate.
Here we have assumed there is no water source or sink in the model
domain
(but there may be water injection or extraction at the boundaries in
order to maintain certain boundary conditions).
Both $K$ and $h$ vary with $x$.
We view $K$ as the ``controlling'' physical attribute and $h$ as the ``outcome''.
Given the 1-D conductivity field $K$,
we can solve this equation to find the head $h$ everywhere in the model
domain.
The scientific question is to infer $K$ given measured $h$ at a number
of locations.

\begin{figure}
\includegraphics[scale=.8]{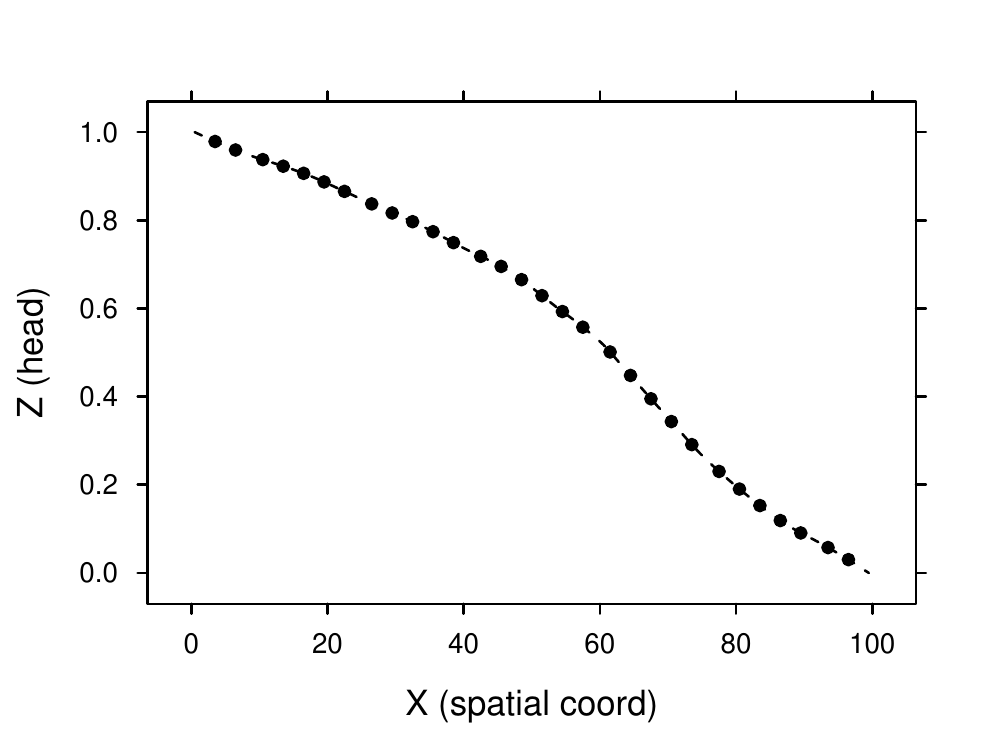}
\caption{Steady-state head level (m)
    in the synthetic log-conductivity field,
    under specified boundary conditions,
    of example~1 in section~\ref{sec:examples}.
    The dots indicate synthetic head measurements.}
\label{fig:darcy-forward}
\end{figure}

Suppose $h$ at 30 locations in the field
were available as forward data.
The forward model $\M$ involved transforming
$\Y$ to $K = \rme^\Y$, solving~(\ref{eq:ex1-forward}) for $h$
under specified  boundary conditions
(that is, $h = 1$ at the left boundary and $h=0$ at the right boundary),
and extracting the 30 values of $h$ at the measurement locations.
The synthetic forward data (figure~\ref{fig:darcy-forward})
are the output of the forward model given the synthetic field
as input.

Noticing in~(\ref{eq:ex1-forward})
that a scaling of $K$ (or a shifting of $Y$) does not change the
resultant $h$, we need at least one direct measurement of $Y$.
The synthetic $Y$ value at a randomly picked location,
marked in figure~\ref{fig:darcy-field},
was made available to the inversion as linear data.

\begin{figure}
\includegraphics[scale=.8]{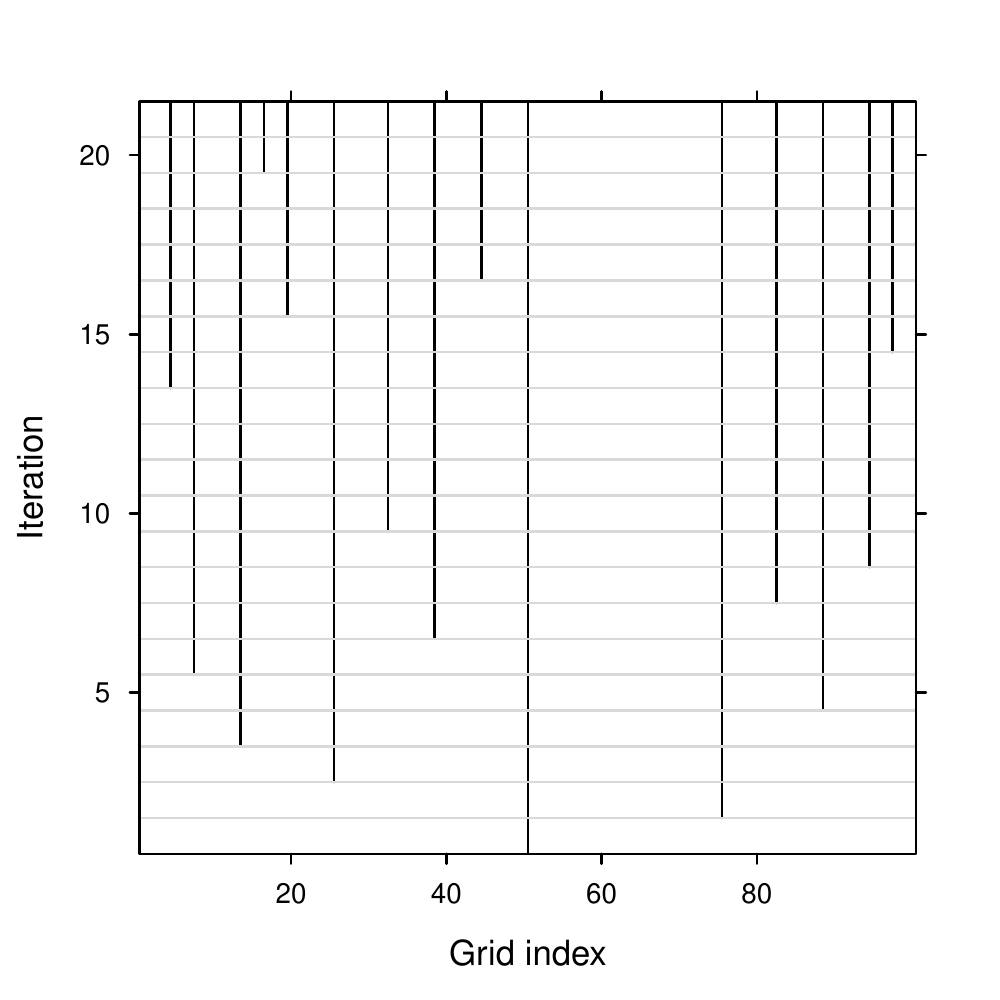}
\caption{Evolution of anchorsets in the iterative algorithm for
    example~1 in section~\ref{sec:examples}.
    In each iteration, the model domain is divided by
    the vertical bars; the average value of $Y$ in each division
    constitutes an anchor.}
\label{fig:darcy-anchor-layout}
\end{figure}

Some summaries and diagnostics of the result are listed
in table~\ref{tab:ex1}.
In the table one sees decisive
increase in $L^*$ and decrease in mad ratios, as desired.
Overall, the number of anchors started at 2 and increased to 16.
Because the geostatistical parameter had 4 components,
the dimension of the model parameter $\anchor$
started at 6 and increased to 20.
The adaptive nature of anchor selection in the algorithm
is vividly depicted in
figure~\ref{fig:darcy-anchor-layout},
which shows the evolution of the anchorsets in the iterations.

\begin{table}
\caption{Summaries and diagnostics for Example~1.}
\label{tab:ex1}
\begin{center}
\begin{tabular}{@{}rrrrrr}
\br
Iteration & Sample size & Anchors
    & $L^*$ & Median $r_{\text{mad}}^{(i)[j]}$
    & Max $r_{\text{mad}}^{(i)[j]}$ \\
\mr
1  & 2400  &  2 & -48.6  & 1.0000  & 1.000 \\
2  & 1950  &  3 &  -7.4  & 0.1829  & 0.346  \\
3  & 1612  &  4 &  31.7  & 0.0354  & 0.188  \\
4  & 1359  &  5 &  35.2  & 0.0402  & 0.146  \\
5  & 1170  &  6 &  43.3  & 0.0290  & 0.138  \\
6  & 1027  &  7 &  49.1  & 0.0270  & 0.086  \\
7  &  920  &  8 &  72.9  & 0.0110  & 0.089  \\
8  &  840  &  9 &  85.1  & 0.0081  & 0.063  \\
9  &  780  & 10 &  84.0  & 0.0095  & 0.034  \\
10 &  735  & 11 &  83.6  & 0.0091  & 0.034  \\
11 &  701  & 11 &  70.0  & 0.0086  & 0.040  \\
12 &  676  & 11 &  85.8  & 0.0096  & 0.025  \\
13 &  657  & 11 &  90.4  & 0.0063  & 0.026  \\
14 &  643  & 12 &  96.0  & 0.0064  & 0.022  \\
15 &  632  & 13 &  99.2  & 0.0051  & 0.023  \\
16 &  624  & 14 &  98.8  & 0.0052  & 0.021  \\
17 &  618  & 15 & 103.3  & 0.0040  & 0.036  \\
18 &  614  & 15 & 106.0  & 0.0037  & 0.037  \\
19 &  610  & 15 & 101.5  & 0.0038  & 0.038  \\
20 &  608  & 16 &  97.8  & 0.0037  & 0.038  \\
\br
\end{tabular}
\end{center}
\end{table}

\begin{figure}
\centering
\includegraphics[scale=.9]{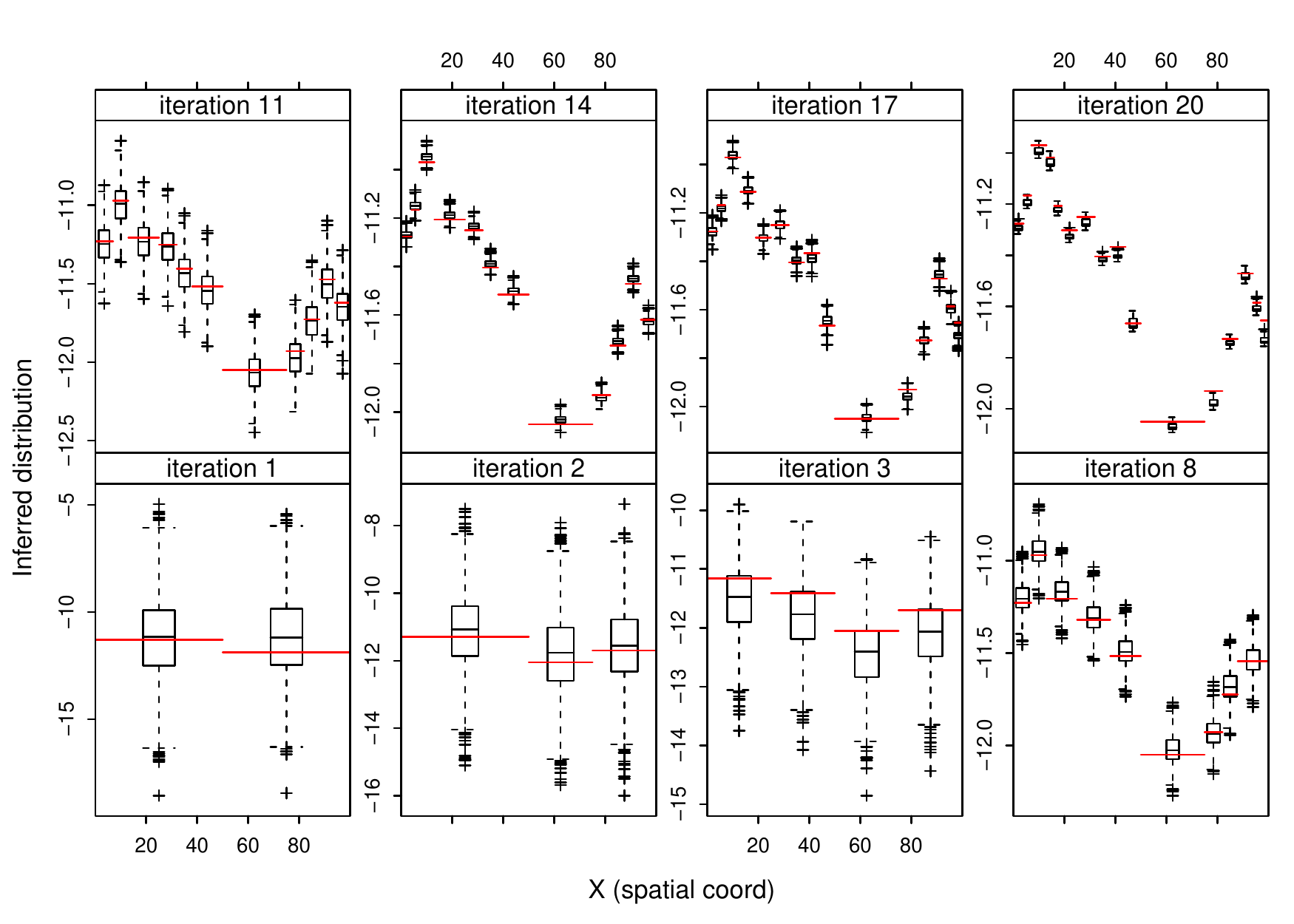}
\caption{Marginal distribution of each anchor,
    represented by a boxplot,
    in the iterative algorithm for
    example~1 in section~\ref{sec:examples}.
    Each box is located horizontally at the center of
    the ``support'' of its corresponding anchor.
    The red lines indicate the ``target value'' of each anchor
    (obtained using the synthetic true field).}
\label{fig:darcy-anchor-bw}
\end{figure}

Figure~\ref{fig:darcy-anchor-bw} shows the inferred marginal distribution
of each anchor, as represented by a boxplot
of the sample obtained in step~1 of the algorithm.
Since the anchorsets evolved in iterations,
we do not see the ``convergence'' of the distribution of a fixed anchorset.
Because the true field is known,
each anchor has a ``target value'' defined by the true $Y$ values.
These target values are indicated in figure~\ref{fig:darcy-anchor-bw}.
It is seen that the estimated distributions homed in on the target
values convincingly after a few iterations.
This strength of convergence for a reasonably large
number of anchors suggests that the parameterization
provides a good presentation of the field.

\begin{figure}
\centering
\includegraphics[scale=.9]{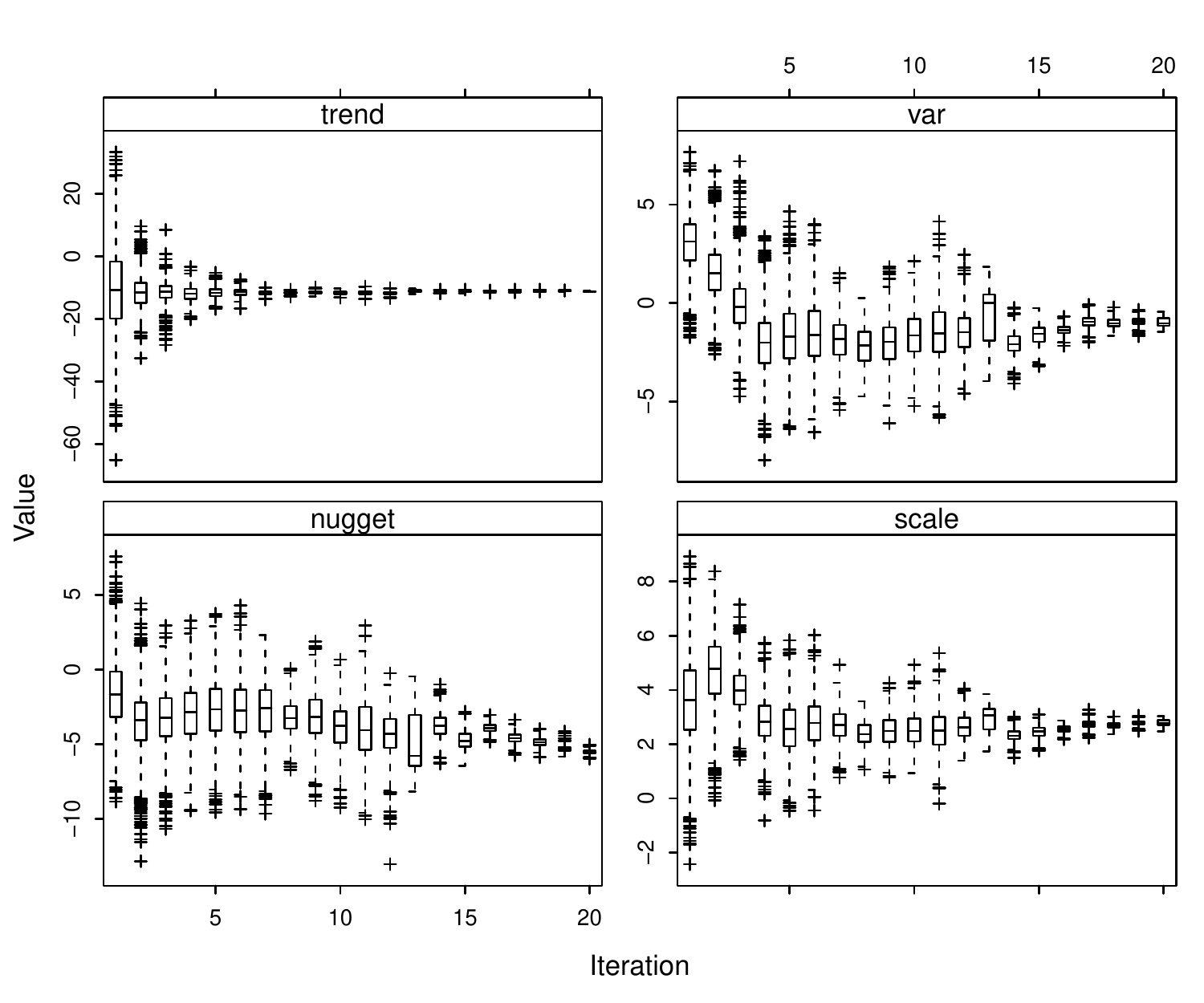}
\caption{Evolution of the estimated marginal posterior distributions
    of each geostatistical parameter in the iterative algorithm for
    example~1 in section~\ref{sec:examples}.
    The distributions are represented by boxplots of the samples
    obtained in step~1 of the algorithm.}
\label{fig:darcy-geost-sample}
\end{figure}

The estimated marginal distributions of the geostatistical parameters
are shown in figure~\ref{fig:darcy-geost-sample}.
It should be stressed that the meaning of the geostatistical parameters
changes during the iterations whenever the anchorset changes.
In addition, interactions between the geostatistical parameters
cause difficulty in their inference (see \citet{Zhang:2012:IPI}).
Compared to the anchor parameters,
the role of geostatistical parameters is more of an intermediate
modeling device.

\begin{figure}
\centering
\includegraphics[scale=.8]{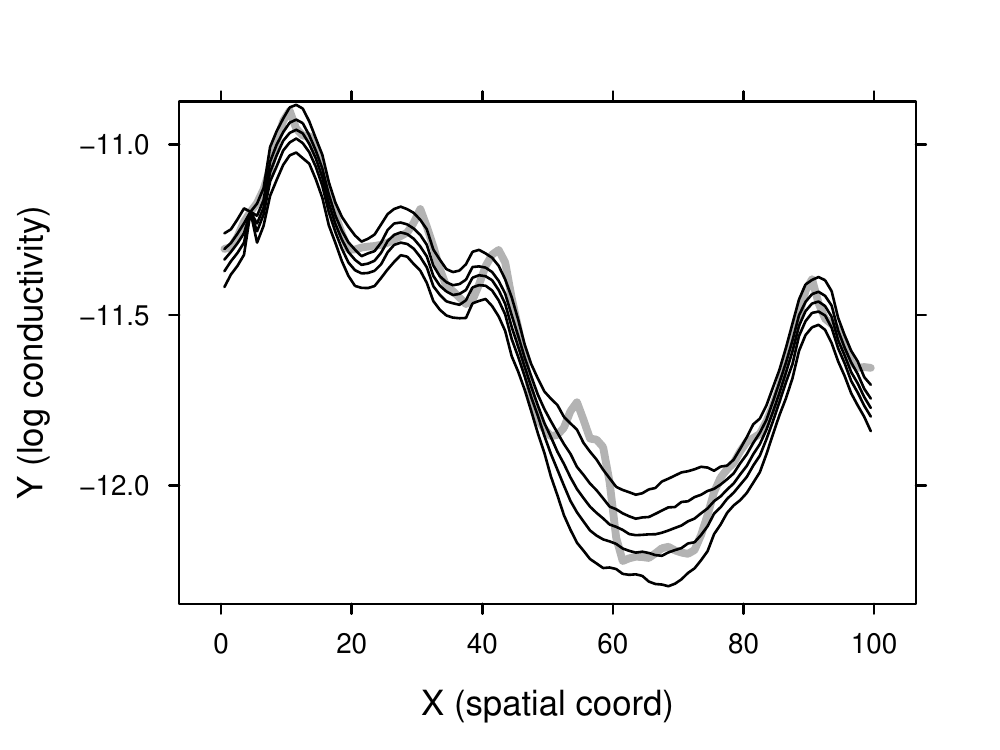}
\caption{Pointwise 5th, 25th, 50th, 75th, and 95th percentiles
    of 1000 field simulations based on the anchorset
    and approximate posterior distribution in the final iteration
    of example~1 in section~\ref{sec:examples}.
    The thick gray curve is the synthetic field.
    (Note how the direct measurement 
    was reproduced exactly.)}
\label{fig:darcy-quantile}
\end{figure}

Recall that the goal of anchored inversion is to characterize the field $Y$.
We created 1000 field realizations based on the anchorset
and approximate posterior distribution in the final iteration
(\ie $f^{(20)}(\anchor)$ in this particular example).
The 5th, 25th, 50th, 75th, and 95th percentiles of the realizations
for each grid point are plotted in figure~\ref{fig:darcy-quantile},
overlaid with the synthetic field for comparison.
The comparison confirms the inversion worked well.

\begin{figure}
\centering
\includegraphics[scale=.9]{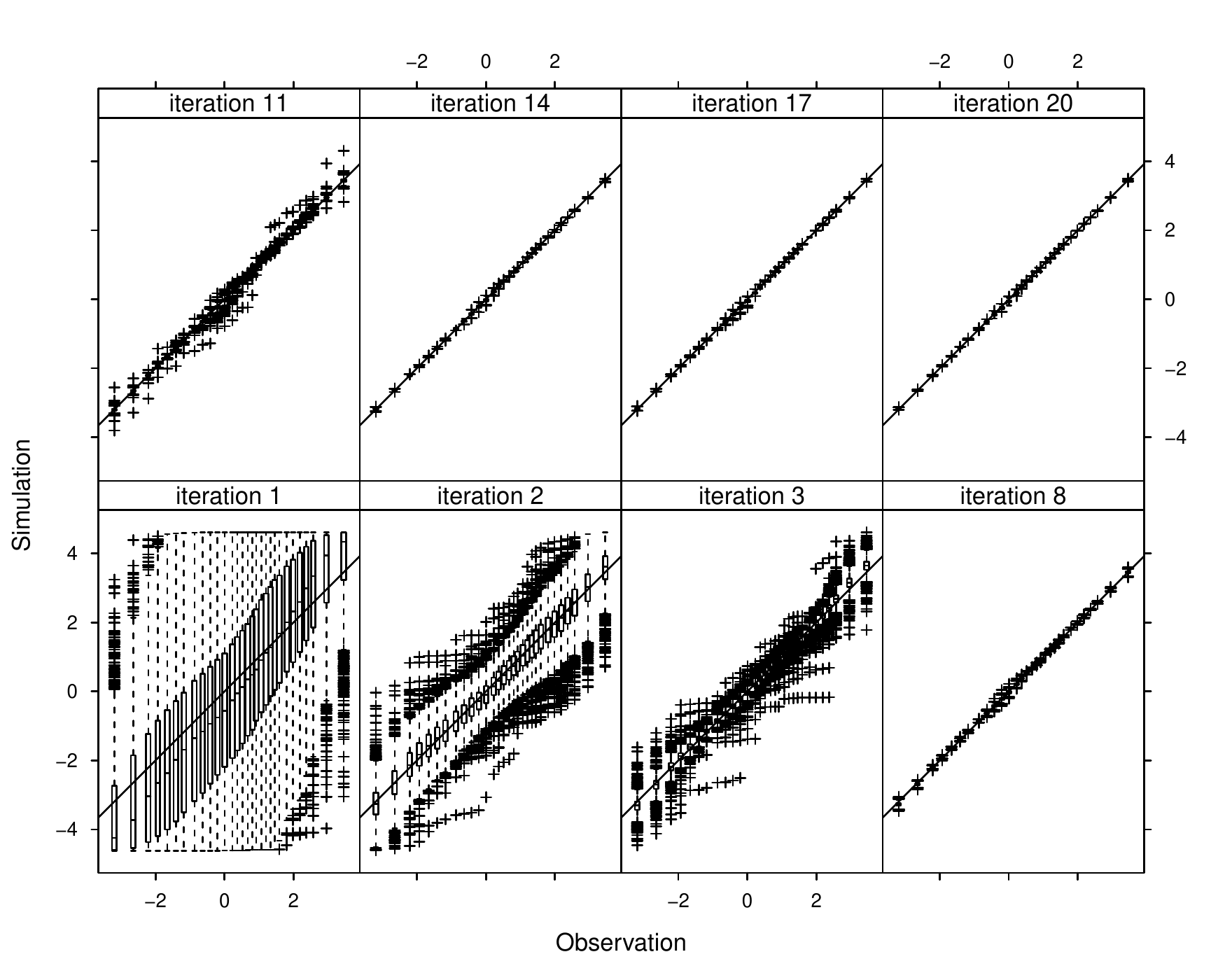}
\caption{Boxplot of each component of the forward predictions
    ($z_i$) sampled in step~2 of selected iterations
    in example~1 of section~\ref{sec:examples}.
    The horizontal location of each boxplot is at
    the observed value of the corresponding forward datum.
    The straight reference lines are $y=x$.}
\label{fig:darcy-pred-by-iter}
\end{figure}

In real-world applications,
we do not know the true field,
therefore can neither compare the field simulations with the truth
nor compare the anchors with their target values.
Perhaps the most concrete assessment is provided by
the capabilities of simulated fields to ``reproduce''
the forward data.
In each iteration of the algorithm,
the sample $\{z_i\}_1^n$ obtained in step~2 represents
``predictions'' of the observation based on
the posterior distribution estimated in the previous iteration.
These predictions are compared with the actual observations
in figure~\ref{fig:darcy-pred-by-iter}.
In each panel (for one iteration),
a vertical box-plot for each component of the $Z$ predictions is drawn
on the horizontal location where the observed value is.
Therefore,
if the predictions reproduce the observations,
the box-plots should fall centered on the $y=x$ line.
The figure demonstrates that
the reproduction starts severely biased and outspread
in the initial iteration and,
after steady adjustment of the mean and reduction of the variance
in subsequent iterations,
achieves precise match with the actual observation
in the final iterations.

\subsection{Example 2: rainfall and surface runoff}
\label{sec:example-2}

Runoff over the land surface generated during and after a rainfall
event is affected by the hydraulic roughness of the land surface
\citep[chap~6]{Vieux:2004:DHM}.
By affecting runoff,
the roughness
also exerts important influence on infiltration, erosion,
and vegetative growth.
However, detailed measurement of the roughness coefficient in a
watershed is impractical.
In many hydrological and agricultural models,
surface roughness is treated as a constant or as
a categorical variable taking a few values based on the type of the land
cover. 
\citet{Huang:2009:ISH} investigate the effect of spatially
heterogeneous roughness on the runoff process.
They consider a one-dimensional overland plane model described by the
following equation system:
\begin{equation}\label{eq:ex2-forward}
\frac{\partial h}{\partial t} + \frac{\partial q}{\partial x} = b,\qquad
s_f = s_0 - \frac{\partial h}{\partial x},\qquad
q = hu,\qquad
u = r^{-1} h^{2/3} s_f^{1/2}
\end{equation}
where
$x$ is the spatial coordinate (m),
$t$ is the time (s),
$h(x, t)$ is the flow depth (m),
$q(x, t)$ is the flow discharge per unit width (\unit{m^2 s^{-1}}),
$b(x, t)$ is the effective rainfall (\unit{m s^{-1}}),
$s_f(x,t)$ is the friction slope (m/m),
$s_0(x)$ is the bed slope (m/m),
$u(x, t)$ is the flow velocity (\unit{m s^{-1}}),
and $r(x)$ is the Manning roughness coefficient
    (\unit{s m^{-1/3}}).
In this system,
the rainfall input $b$ and the bed slope $s_0$ are considered known.
In addition,
the initial condition $h(x, 0) = q(x, 0) = 0$
and boundary condition $q(0, t) = 0$ are assumed.
These conditions state that the surface is dry at the beginning of the
rainfall event, and there is no inflow at the upstream boundary.

\begin{figure}
\includegraphics[scale=.8]{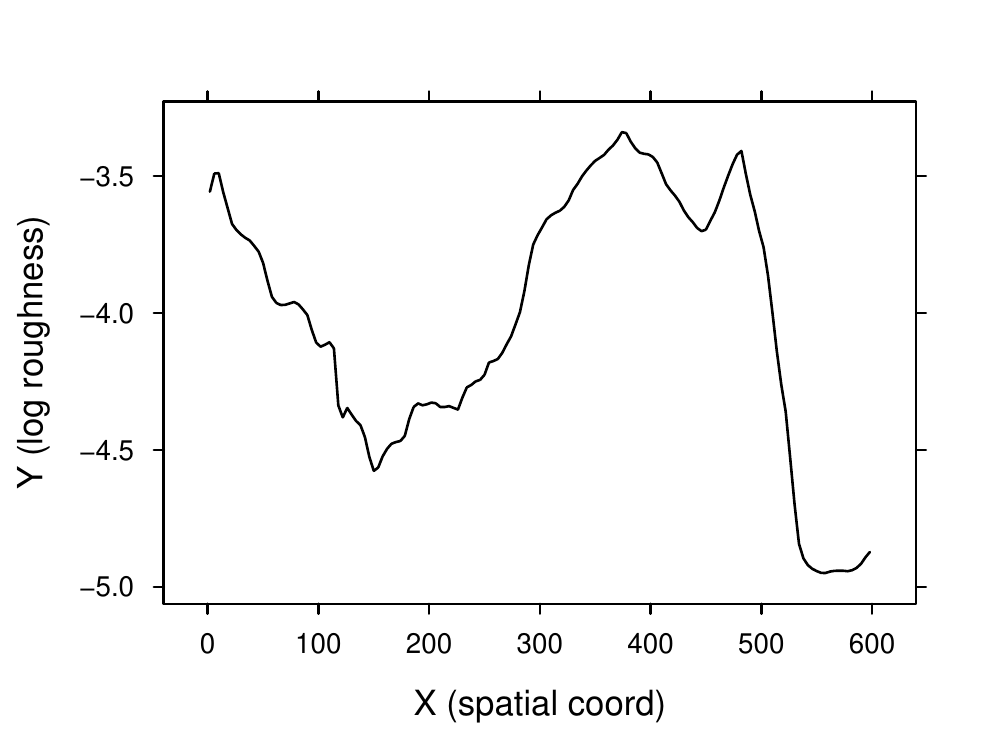}
\caption{Synthetic field of log roughness
    (\unit{s m^{-1/3}}) for example~2
    in section~\ref{sec:examples}.}
\label{fig:runoff-field}
\end{figure}

Whereas \citet{Huang:2009:ISH} investigate
the resultant $h$ and $q$ under various configurations of
synthetic roughness $r(x)$,
we considered the inverse problem:
infer the spatial field of $r(x)$ given observations of $h$ and $q$.
The synthetic log-roughness field we used is shown in
figure~\ref{fig:runoff-field}.
The field consists of 150 grid points.

\begin{figure}
\includegraphics[scale=.8]{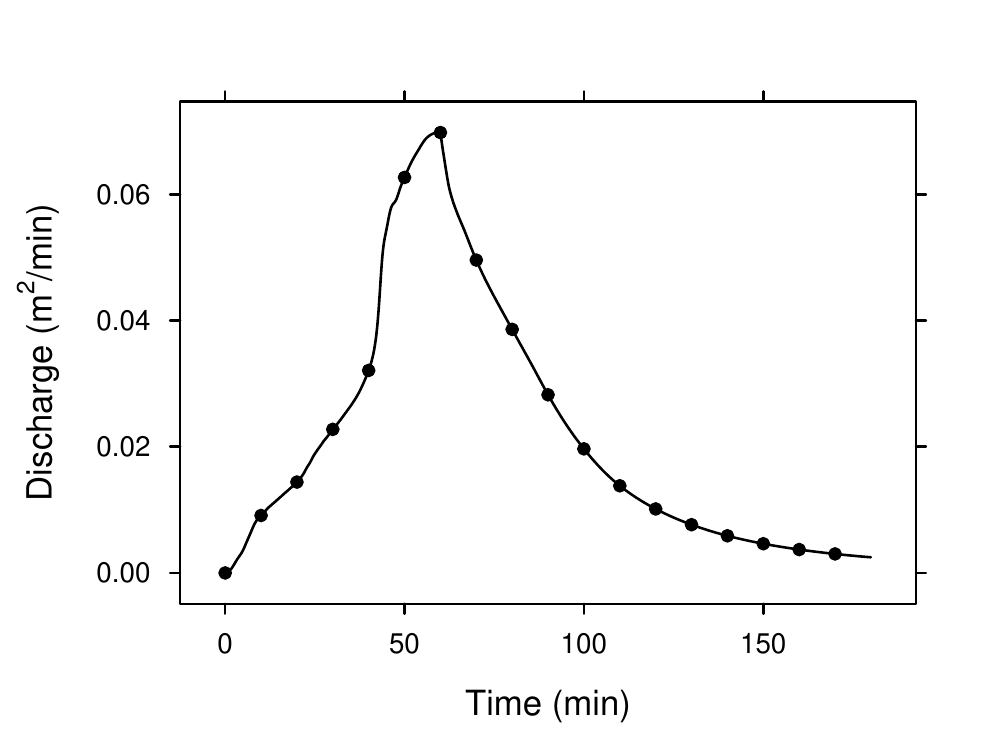}
\includegraphics[scale=.8]{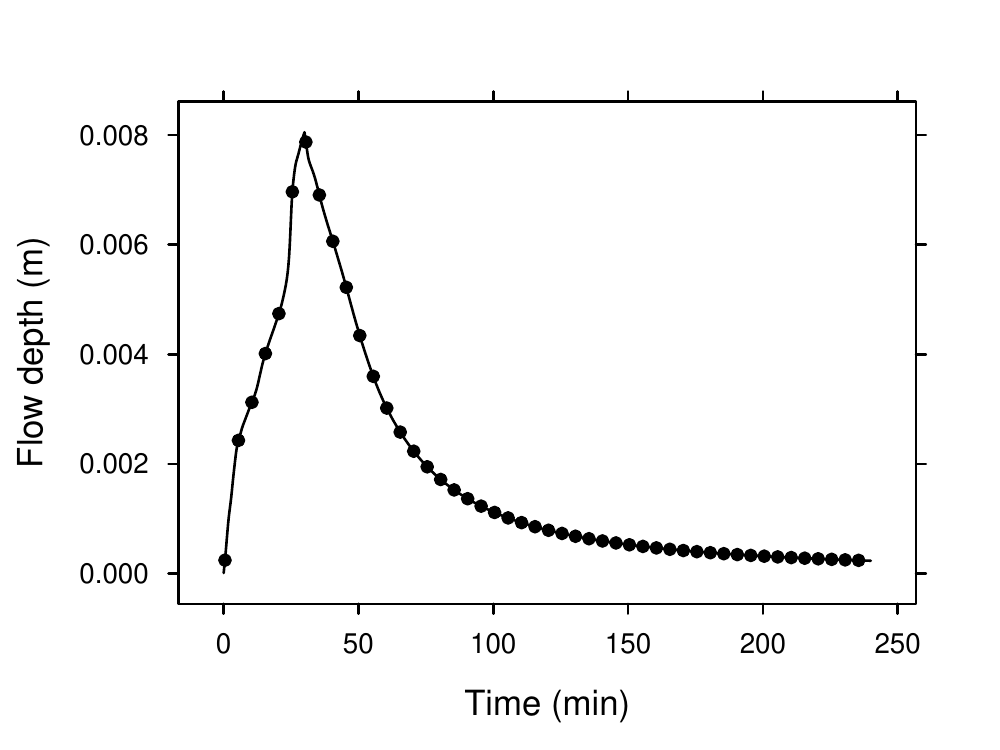}
\caption{Discharge (top; in the first rainfall and measurement scenario)
    and flowdepth (bottom; in the second rainfall and measurement scenario)
    processes at the downstream boundary in example~2
    of section~\ref{sec:examples}.
    The dots indicate synthetic measurements.}
\label{fig:runoff-forward}
\end{figure}

To make it more interesting,
we constructed two synthetic rainfall events and corresponding observation scenarios.
In the first event,
rainfall lasted for 60 minutes with a uniform intensity of
5\unit{mm/hr}.
The discharge at the downstream boundary
was measured every 10 minutes for 3 hours
(that is, during the rain and in the subsequent 120 minutes).
In the second event,
rainfall lasted for 30~minutes with a uniform intensity of
20\unit{mm/hr}.
The flowdepth at the downstream boundary
was measured every 5 minutes for 4 hours
(that is, during the rain and in the subsequent 210 minutes).
These two time-series datasets were combined to form a forward data
vector of length 66.
The discharge and flowdepth processes at the downstream boundary
in the two rain events,
computed based on the synthetic roughness field,
are depicted in figure~\ref{fig:runoff-forward}.
We took $\Y = \log r$ and modeled $\Y$ as a Gaussian process.
The forward function $\M$ involved
obtaining $r = \rme^{\Y}$,
solving the equation system~(\ref{eq:ex2-forward}) for $h$ and $q$
based on each of the two rainfall scenarios,
extracting $q$ (or $h$) at the specified space and time coordinates,
and taking the combined vector $(\log q, \log h)$ as the forward data $\z$.
Note here the forward data contained some discharge measurements and some
flowdepth measurements.
These two physical quantities are of incomparable natures.

In the iterative algorithm,
principal component analysis reduced the dimension
of the forward data from 66 down to 9--19.
Such significant dimension reduction was expected
given the high level of auto-correlation in the time series data.

\begin{figure}
\includegraphics[scale=.8]{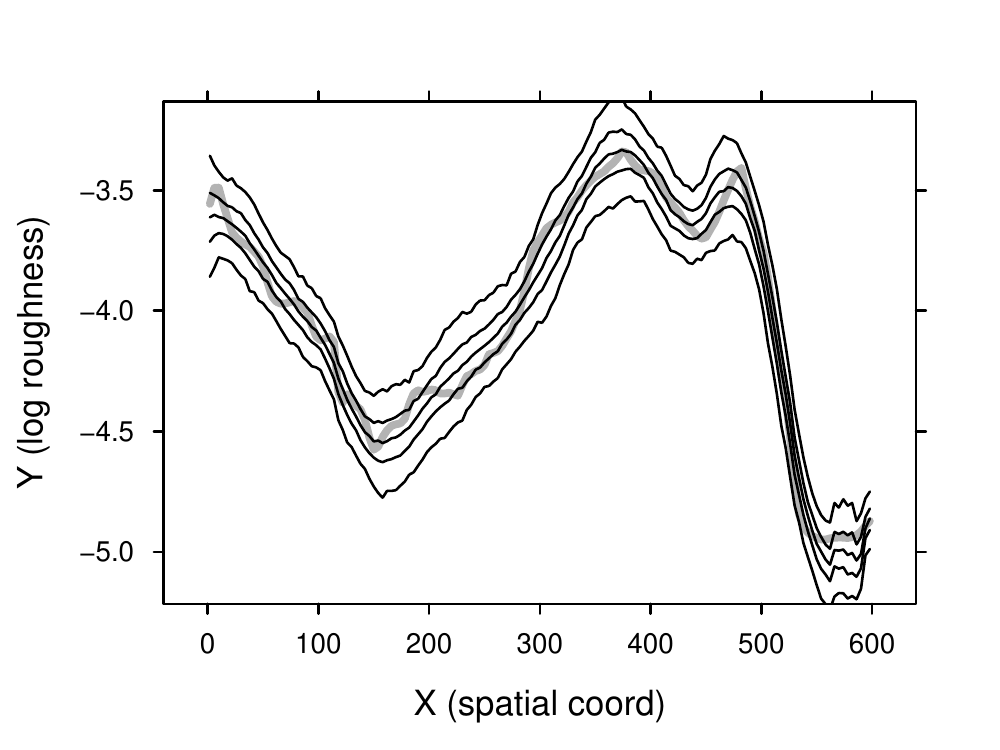}
\caption{Pointwise percentiles of 1000 field simulations
    in example~2 of section~\ref{sec:examples}.
    See the caption of figure~\ref{fig:darcy-quantile} for explanations.}
\label{fig:runoff-quantile}
\end{figure}

Performance of the inversion is demonstrated by 1000 field realizations
created based on the final approximation to the posterior distribution
of $\anchor$.
Representative percentiles of the simulated realizations
are shown in figure~\ref{fig:runoff-quantile}
with comparison to the synthetic field,
confirming the potential of the inversion method.

\subsection{Example 3: seismic tomography}
\label{sec:example-3}

The propagation of natural or man-made seismic waves through the earth
is controlled by certain physical properties of the rock media
along the path of the wave propagation.
It is conventional to think of the media as constituting a ``velocity
field'', that is, each location is characterized by a velocity which is
determined by relevant rock properties at that location.
Suppose seismic P-wave emanates from a source location
and is detected at a receiver location, where the ``first-arrival
traveltime'' of the wave is recorded.
The ray path (\ie route of the wave propagation)
is a complicated function of the velocity field;
in particular, it is not a straight line
where the velocity field is not uniform.
The traveltime is determined by the velocities on the ray path
from the source to the receiver.
If one knows the velocity field,
then both the ray path and the traveltime can be calculated
\citep{Vidale:1988:FDC, Vidale:1990:FDC}.
The inverse problem,
namely inference of the velocity field given measurements of traveltimes,
is called ``seismic tomography''.
First developed in medical imaging
\citep{Arridge:1999:OTM},
the tomography technique has seen wide applications in fields such as
geophysics \citep{Nolet:2008:BST}
and hydrology \citep{Yeh:2000:HTD}.

\begin{figure}
\centering
\includegraphics[scale=.9]{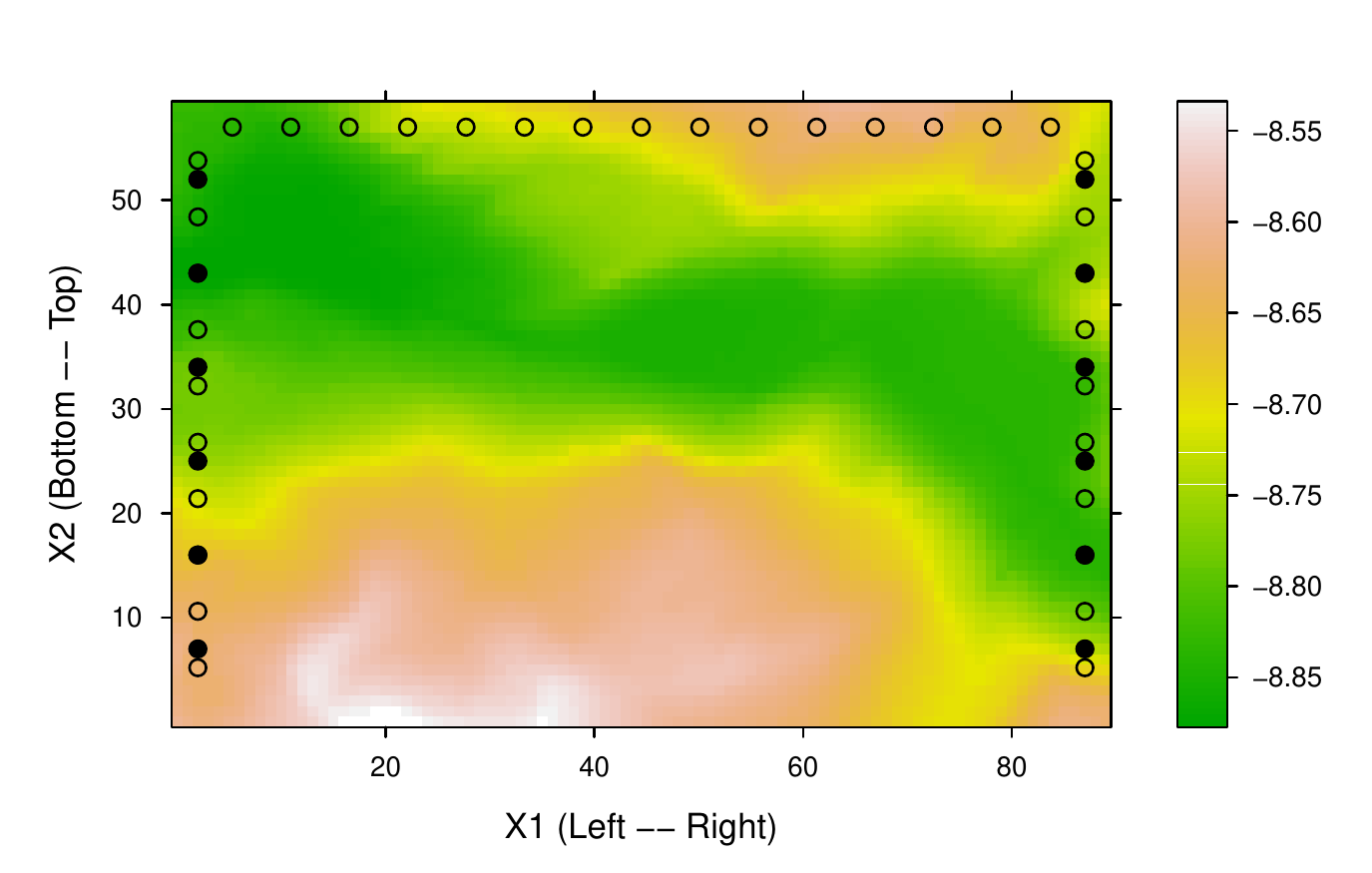}
\caption{Synthetic field of log slowness (second/meter)
    in example~3 of section~\ref{sec:examples}.
    The dots indicate wave sources; the circles indicate receivers.}
\label{fig:traveltime-field}
\end{figure}

We constructed a synthetic tomography problem in two-dimensional space.
Define ``slowness'' $s$ (\unit{s m^{-1}}) as the reciprocal of velocity.
Figure~\ref{fig:traveltime-field} shows a synthetic log-slowness
($Y$)
field (a vertical profile dipping down the ground surface),
discretized into a $60\times 40$ grid.
Suppose we conducted a tomography experiment
with 6 sources and 10 receivers on each vertical boundary,
and 15 receivers on the top boundary.
Waves from each source were detected by the receivers on the opposite
vertical boundary and on the top boundary.
The sources and receivers are marked in figure~\ref{fig:traveltime-field}.
Such a layout of sources and receivers is typical of
``crosshole tomography'' \citep[sec.~2.1]{Ivansson:1987:CTT}.

\begin{figure}
\centering
\includegraphics[scale=.9]{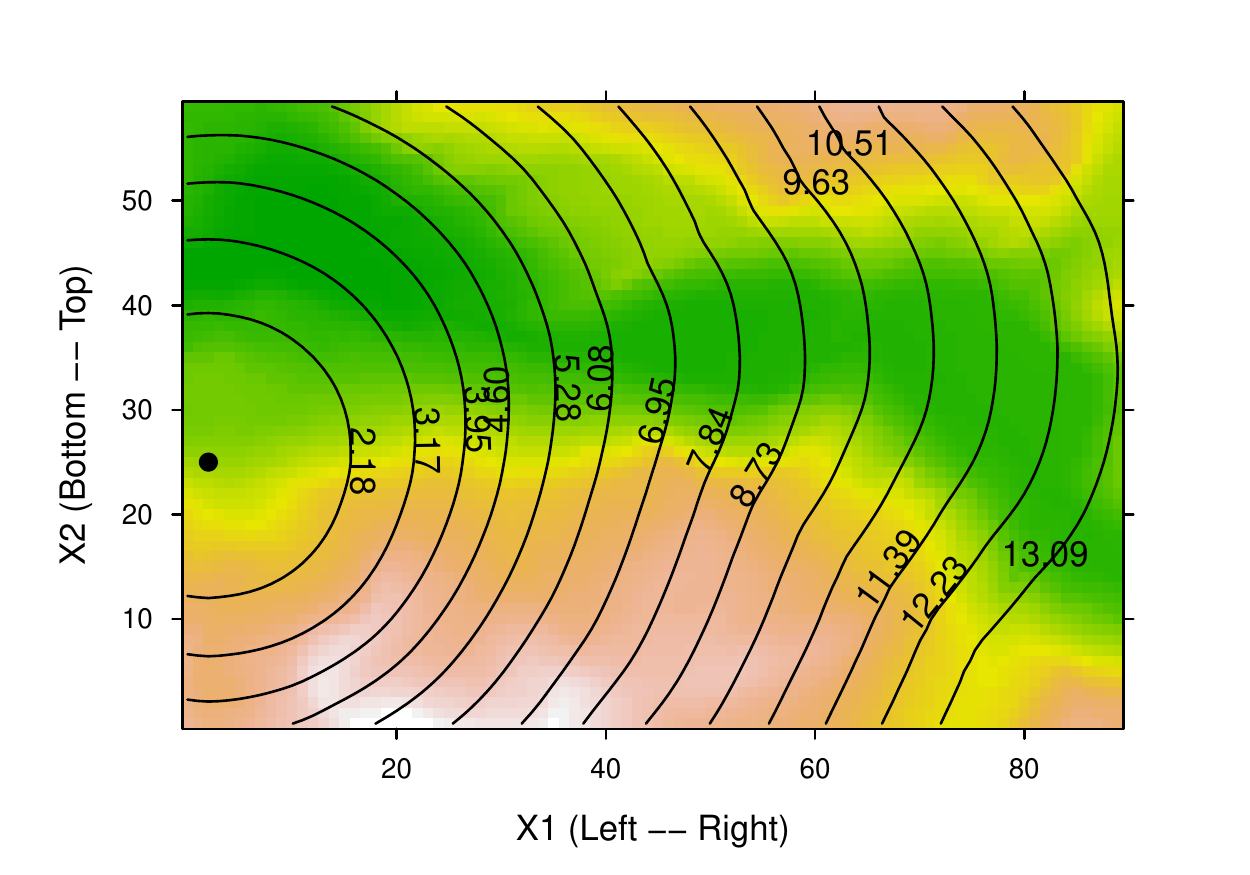}
\caption{First-arrival times ($10^{-3}\unit{s}$) of seismic wave,
    originating from the source indicated by the dot on the left,
    in the synthetic slowness field (on $60\times 40$ grid)
    of example~3 in section~\ref{sec:examples}.}
\label{fig:traveltime-forward}
\end{figure}

The wave propagation is described approximately by the
Eikonal equation \citep[Appendix~C]{Shearer:2009:IS}:
\begin{equation}\label{eq:ex3-eikonal}
\biggl(\frac{\partial\, t(x_1, x_2)}{\partial x_1}\biggr)^2
+ \biggl(\frac{\partial\, t(x_1, x_2)}{\partial x_2}\biggr)^2
= \bigl(s(x_1, x_2)\bigr)^2
,
\end{equation}
where $x_1$, $x_2$ are the spatial coordinates,
$t(x_1, x_2)$ is the traveltime from a designated source
to $(x_1, x_2)$, and $s(x_1, x_2)$ is the slowness at location
$(x_1, x_2)$.
To illustrate a solution of this system,
the contours of first-arrival traveltimes in the synthetic field
from a source located on the left boundary
are shown in figure~\ref{fig:traveltime-forward}.
This figure clearly illustrates the curved paths of the seismic wave
in the heterogeneous media.
Note that the wave travels faster in regions of lower slowness
(logically enough!).

Solving~(\ref{eq:ex3-eikonal}) in the synthetic slowness field
and taking logarithmic transform of the resultant traveltimes,
we obtained
$2 \times 6 \times (10 + 15) = 300$ synthetic traveltime measurements.
The solution used the algorithm described by
\citet{Vidale:1988:FDC, Vidale:1990:FDC}.

To reduce the computational cost,
we aggregated the synthetic field into a $30\times 20$ grid,
that is, a fourth of the resolution of the synthetic field.
This introduces \emph{model error}: the synthetic forward model
and the forward model used in inversion are not identical---they use
different numerical grids.
This error 
is directly analogous to what discretization does to
the conceptually continuous field in real-world applications.
It may not be trivial to characterize this error quantitatively.
In this example we chose to ignore this model error.
Although in this particular case the error is small,
the notion of ignoring the error is important in illustrating the
principles of the anchored inversion methodology.
In contrast, some other methods rely on the notion of ``error''
to construct regularization criteria or likelihood functions
(see section~\ref{sec:anchor-motivation}).

\begin{figure}
\centering
\includegraphics[scale=.9]{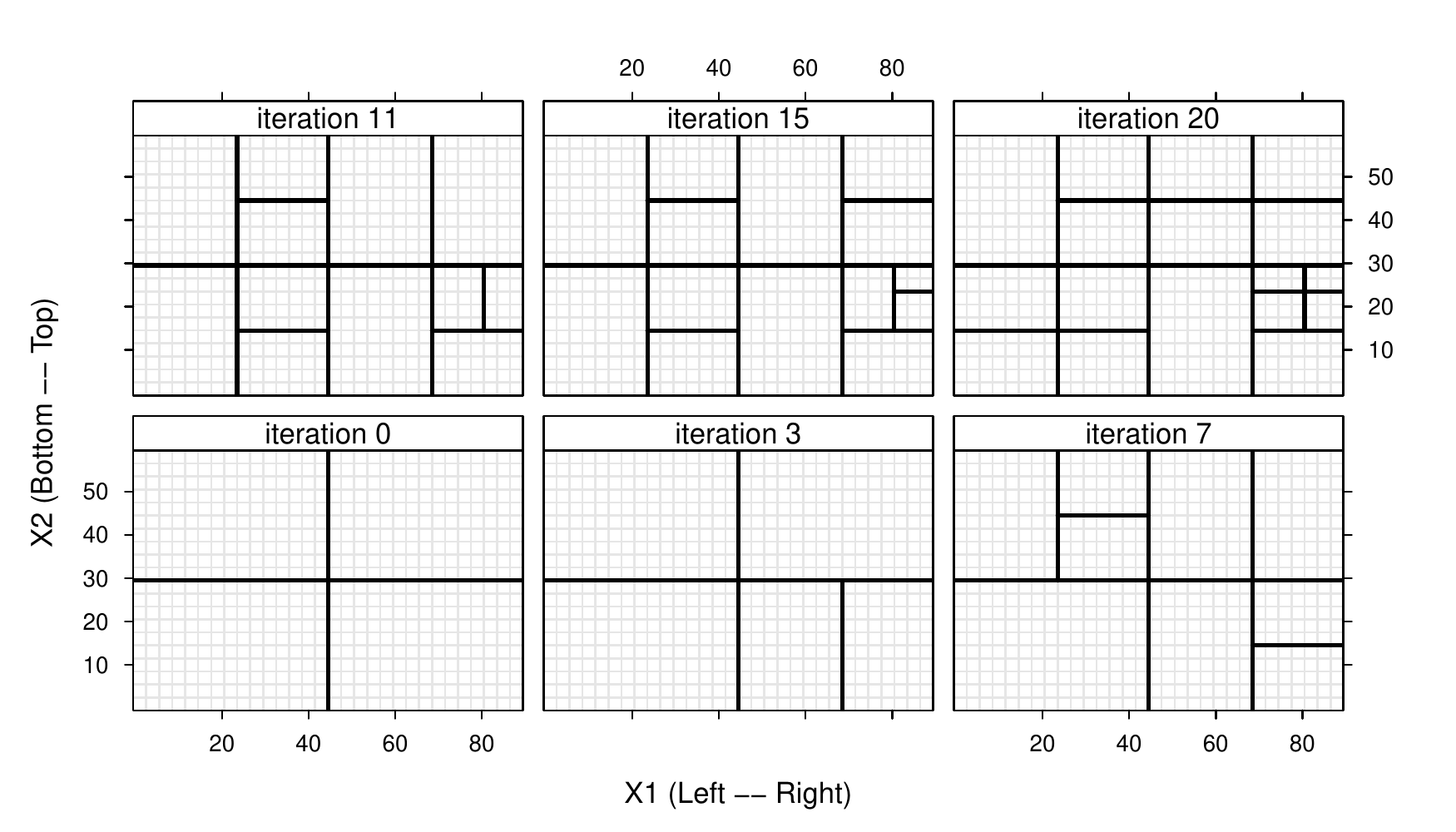}
\caption{Layout of anchorsets in selected iterations
    of example~3 in section~\ref{sec:examples}.
    The model domain is divided into sub-domains;
    the average value of $Y$ in each sub-domain constitutes
    an anchor.}
\label{fig:traveltime-anchor-layout}
\end{figure}

In the iterations of the algorithm,
the dimension of the forward data was reduced from 300 to 25--107,
thanks to the principal component analysis in step~3 of the algorithm.
The algorithm started with 4 anchors, dividing the model domain
into $2\times 2$ sub-domains.
The parameterization evolved in the 20 iterations and
arrived at an anchorset with 16 anchors,
as depicted in figure~\ref{fig:traveltime-anchor-layout},
which once again illustrates the adaptive nature of the algorithm
in terms of automatically selecting an anchorset.

\begin{figure}
\centering
\includegraphics[scale=.9]{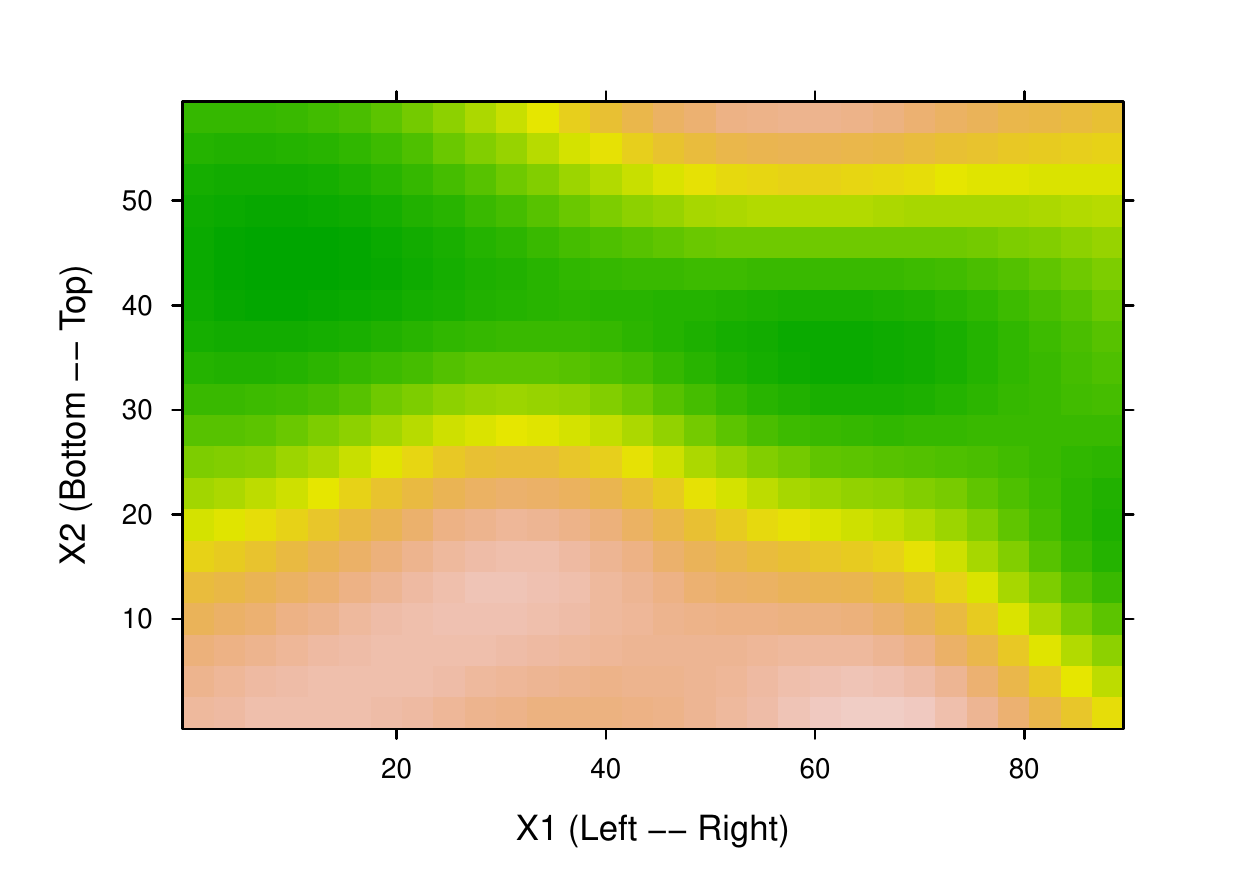}
\caption{Pointwise median of 1000 field simulations
    based on the anchorset and approximate posterior distribution
    in the final iteration of example~3 in section~\ref{sec:examples}.}
\label{fig:traveltime-median}
\end{figure}

Based on the final approximation to the posterior distribution of the
model parameter $\anchor$,
we created 1000 field realizations (on the $30\times 20$ grid).
The point-wise median of these realizations is shown in
figure~\ref{fig:traveltime-median}.
Comparison with the synthetic field
(aggregated to the $30\times 20$ grid) confirms
that the inversion captured important features of the field.
The mean of the point-wise standard deviations of the field realizations
is 0.015, whereas the mean of the point-wise median absolute deviations (from
the synthetic truth) is 0.036.
These values should be considered with reference to
the range of the synthetic field, which is $(-8.88, -8.53)$.


\section{Conclusion}
\label{sec:conclusion}

We have proposed a general method for modeling a spatial random field,
as a Gaussian process, that is tailored to making use of
measurements of nonlinear functionals of the field.
These nonlinear functionals are ``forward processes (or models)''
observed in experimental or natural settings.
Inverse of nonlinear forward processes is a major topic with wide
scientific applications, as demonstrated in
sections \ref{sec:intro} and~\ref{sec:examples}.

The method revolves around certain deliberately chosen
linear functionals of the field called ``anchors''.
The anchor parameterization achieves two goals.
First, it reduces the dimensionality of the parameter space
from that of the spatial field vector to that of the anchor vector
(plus possibly a small number of other parameters such as geostatistical
parameters).
As a result, the dimensionality of the statistical inversion problem
is separate from the spatial resolution of the forward model.
Second, the parameterization establishes a statistical connection
between the model parameter (\ie anchors) and the data
(\ie forward process outcome).
This statistical connection is known in mechanism
and can be studied through simulations.
The existence of this connection makes the inversion task a
statistical one, whether or not the data contain model and measurement
errors.
This makes the method a fundamental departure from some existing
approaches.

Although the framework has a Bayesian flavor,
it does not go down a typical Bayesian computational path,
because the likelihood function is unknown.
The computational core lies in an iterative algorithm
(section~\ref{sec:inference}) centered on kernel density estimation.
The algorithm is general and flexible in several aspects.
First, it does not require a statistical analysis of the parameter-data
connection, but only needs the ability to evaluate the forward model
(usually by running a numerical code),
which is treated as a black box operation.
Although this study has focused on deterministic forward models,
it is possible to accommodate stochastic ones.
Second, the algorithm is flexible with regard to the prior
specification.
Third,
the algorithm is able to accommodate flexible forms of
errors (section~\ref{sec:accommodate-errors}).
Fourth,
different forward models (hence forward data) could be used
in different iterations.
The last point, not elaborated in this article,
makes it possible to assimilate multiple datasets sequentially
or in other flexible ways.
These features make
the algorithm a general procedure for approximating the posterior distribution
where the likelihood is unknown but the model (which corresponds to the
likelihood) can be simulated.

A critical component of the method concerns
selecting the anchor parameterization in an adaptive, automatic fashion
as iteration proceeds (section~\ref{sec:choice-of-anchors}).
The resultant anchor configuration is tailored to
features of the field $Y$
and
the computational effort already invested.
For example,
if more iterations are performed,
the algorithm in general will introduce
more anchors to achieve more detailed
representation of the field.

We used several synthetic examples to demonstrate a number of features
of the methodology,
including the capabilities of
using linear data (example~1),
using highly correlated data (all examples, especially examples 2
and~3),
assimilating multiple datasets of different
physical natures (example~2),
and ignoring model or measurement errors (example~3).
In view of the dimensionality of the parameter vector (in the tens)
and that of the data (tens or hundreds),
the algorithm showed potential in efficiency.

We have implemented the method in a \texttt{R} package
named \texttt{AnchoredInversion}.
Currently the package uses the packages
\texttt{RandomFields} \citep{Schlather:2001:SAR}
for simulating random fields
and \texttt{MASS} (its function \texttt{mvrnorm})
\citep{Venables:2002:MASS}
for generating multivariate normal random numbers.
The graphics in this article were created using the
package \texttt{lattice} \citep{Sarkar:2008:LMD}.

\bigskip
{\small%
\textbf{Acknowledgement:}\ \ 
The work presented in this article was in part supported by
the Excellent State Key Lab Fund no.\@ 50823005,
National Natural Science Foundation of China,
and
the R\&D Special Fund for Public Welfare Industry no.\@ 201001080,
Chinese Ministry of Water Resources.
The NED data (used in this study as synthetic data for the examples)
from the National Map Seamless Server
are open to the public, and are available
from U.S.\@ Geological Survey, EROS Data Center, Sioux Falls, South
Dakota.
Professor Paul Switzer suggested presenting the geostatistical
parameterization as an extension (instead of being integrated from the
beginning).
The author thanks an anonymous reviewer for constructive comments.
}


\newpage

\twocolumn

\onecolumn

\twocolumn

\onecolumn

\end{document}